\documentclass[11pt]{article}
\usepackage{graphicx}
\usepackage{amsmath}
\RequirePackage{xspace}

\newcommand{\BABARPubYear}    {02}

\newcommand{\BABARConfNumber} {21}
\newcommand{\SLACPubNumber} {9307}

\usepackage{relsize}
\def\babar{\mbox{\slshape B\kern-0.1em{\smaller A}\kern-0.1em
    B\kern-0.1em{\smaller A\kern-0.2em R}}}
\def\pep2{PEP-II}
\def\Kz {\ensuremath{K^0}\xspace}
\def\Kbar    {\kern 0.18em\overline{\kern -0.18em K}{}\xspace}
\def\Kzbar {\ensuremath{\Kbar^0}\xspace}
\def\Dbar    {\kern 0.18em\overline{\kern -0.18em D}{}\xspace}
\def\Dzbar {\ensuremath{\Dbar^0}\xspace}
\def\Bz {\ensuremath{B^0}\xspace}
\def\Bbar    {\kern 0.18em\overline{\kern -0.18em B}{}\xspace}
\def\Bzbar {\ensuremath{\Bbar^0}\xspace}
\def\BdashBbar {\Bz-\Bzbar}
\def\BzBzbar {\ensuremath{\Bz\Bzbar}\xspace}
\def\Btag {\ensuremath{B_{\rm tag}}\xspace}
\def\tauBz      {\ensuremath{\tau_{B^0}}\xspace}
\def\tauBp      {\ensuremath{\tau_{B^+}}\xspace}
\def\dm  {\ensuremath{\delta m}\xspace}
\def\Dm		{\ensuremath{\Delta m_{d}}\xspace}
\def\Dt		{\ensuremath{\Delta t}\xspace}
\def\sigmaDt {\ensuremath{\sigma_{\Dt}}\xspace}
\def\sigmaz {\ensuremath{\sigma_{z}}\xspace}
\def\Dttrue		{\ensuremath{{\Delta t}_{\rm true}}\xspace}
\def\Dtmeas		{\ensuremath{{\Delta t}_{\rm meas}}\xspace}
\def\dDt {\ensuremath{\delta\Delta t}\xspace}
\def\Dz		{\ensuremath{\Delta z}\xspace}
\def\ks  {\ensuremath{K^0_S}\xspace}
\def\FourS {\ensuremath{\Upsilon(4S)}\xspace}
\def\dz {\ensuremath{D^0}\xspace}
\def\dst {\ensuremath{D^{*}}\xspace}

\def\dstm {\ensuremath{D^{*-}}\xspace}

\def\btodstlnu  {\ensuremath{B^0 \rightarrow D^{*-}\ell^+\nu_\ell}\xspace}
\def\dstlnu  {\ensuremath{D^{*-}\ell^+\nu_\ell}\xspace}
\def\dstl  {\ensuremath{D^{*-}\ell^+}\xspace}
\def\dste  {\ensuremath{D^{*-}e^+}\xspace}
\def\dstmu  {\ensuremath{D^{*-}\mu^+}\xspace}
\def\dsttodpi  {\ensuremath{D^{*-}\rightarrow \Dzbar \pi^-}\xspace}
\def\tdstl {\ensuremath{t_{D^*\ell}}\xspace}
\def\ttag  {\ensuremath{t_{\rm tag}}\xspace}

\def\massdiff  {\ensuremath{m(D^*) - m(D^0)}\xspace}
\def\thby {\ensuremath{\theta_{B,\dst\ell}}\xspace}
\def\thbyfl {\ensuremath{\theta_{B,\dst(-\ell)}}\xspace}
\def\costhby {\ensuremath{\cos\thby}\xspace}
\def\costhbyfl {\ensuremath{\cos\thbyfl}\xspace}
\def\zdstl {\ensuremath{z_{D^*\ell}}\xspace}
\def\ztag {\ensuremath{z_{\rm tag}}\xspace}
\def\bdstl {\ensuremath{B_{D^*\ell}}\xspace}
\def\btag {\ensuremath{B_{\rm tag}}\xspace}
\def\chisq {\ensuremath{\chi^2}\xspace}
\def\lepton {\ensuremath{{\tt lepton}}\xspace}
\def\kaon {\ensuremath{{\tt kaon}}\xspace}
\def\ntone {\ensuremath{{\tt NT1}}\xspace}
\def\nttwo {\ensuremath{{\tt NT2}}\xspace}
\def\ntthree {\ensuremath{{\tt NT3}}\xspace}
\def\wBp {\ensuremath{\omega_{B^+}}\xspace}
\def\wBz {\ensuremath{\omega_{B^0}}\xspace}
\def\fBp {\ensuremath{f_{B^+}}\xspace}

\long\def\inst#1{\par\nobreak\kern 4pt\nobreak
  {\it #1}\par\vskip 10pt plus 3pt minus 3pt}

\begin{document}
{\pagestyle{empty}

\begin{flushright}
\babar-CONF-\BABARPubYear/\BABARConfNumber \\
SLAC-PUB-\SLACPubNumber \\
July 2002 \\
\end{flushright}

\par\vskip 2cm

\begin{center}
\Large \bf Simultaneous Measurement of the $B^0$ Meson Lifetime and Mixing Frequency 
with $B^0 \rightarrow D^{*-}\ell^+\nu_\ell$ Decays
\end{center}
\bigskip

\begin{center}
\large The \babar\ Collaboration\\
\mbox{ }\\
July 24, 2002
\end{center}
\bigskip \bigskip

\begin{center}
\large \bf Abstract
\end{center}
We measure the $B^0$ lifetime $\tauBz$ and 
the \BdashBbar oscillation frequency $\Dm$
with a sample of approximately 14,000 exclusively reconstructed
$B^0 \rightarrow D^{*-}\ell^+\nu_\ell$ signal events, 
selected from 23 million $B\overline B$ pairs recorded
at the $\Upsilon(4S)$ resonance with the \babar\ detector at
the Stanford Linear Accelerator Center. 
The $b$-quark flavor of the other $B$ at the time of decay and 
its decay position are determined inclusively.
The lifetime and oscillation frequency are measured 
simultaneously with an unbinned maximum-likelihood fit that uses, 
for each event, the measured difference in $B$ decay times ($\Dt$), 
the calculated uncertainty on $\Dt$,
the signal and background probabilities,
and $b$-quark tagging information for the other $B$.
The preliminary results are 
$$\tauBz = (1.523^{+0.024}_{-0.023} \pm 0.022)~\rm{ps}$$
and
$$\Dm = (0.492 \pm 0.018 \pm 0.013)~\rm{ps}^{-1}.$$
The statistical correlation coefficient between \tauBz\ and \Dm\ is $-0.22$.

\vfill
\begin{center}
Contributed to the 31$^{st}$ International Conference on High Energy Physics,\\ 
7/24---7/31/2002, Amsterdam, The Netherlands
\end{center}

\vspace{1.0cm}
\begin{center}
{\em Stanford Linear Accelerator Center, Stanford University, 
Stanford, CA 94309} \\ \vspace{0.1cm}\hrule\vspace{0.1cm}
Work supported in part by Department of Energy contract DE-AC03-76SF00515.
\end{center}

\newpage
} 


\begin{center}
\small

The \babar\ Collaboration,
\bigskip

B.~Aubert,
D.~Boutigny,
J.-M.~Gaillard,
A.~Hicheur,
Y.~Karyotakis,
J.~P.~Lees,
P.~Robbe,
V.~Tisserand,
A.~Zghiche
\inst{Laboratoire de Physique des Particules, F-74941 Annecy-le-Vieux, France }
A.~Palano,
A.~Pompili
\inst{Universit\`a di Bari, Dipartimento di Fisica and INFN, I-70126 Bari, Italy }
% ...

\end{center}\newpage

\setcounter{footnote}{0}


\section{Introduction and analysis overview}

\label{sec:intro}

The time evolution of $B^0$ mesons is governed by the overall
decay rate $1/\tauBz$ and the \BdashBbar oscillation frequency
\Dm.
The phenomenon of particle-antiparticle oscillations or ``mixing'' 
has been observed in neutral mesons containing a down quark 
and a strange quark ($K$ mesons) or a bottom quark 
($B$ mesons)~\cite{ref:Bmix}.
In the Standard Model of particle physics, $B$ mixing is the result 
of second-order charged weak interactions involving box diagrams 
containing virtual quarks with charge $2/3$.
In $B$ mixing, the diagram containing the top quark dominates.
Therefore, the mixing frequency $\Dm$ is sensitive to the 
Cabibbo-Kobayashi-Maskawa quark-mixing matrix element 
$V_{td}$~\cite{ref:CKM}.
In the neutral $K$ meson system, mixing also has contributions from
real intermediate states accessible to both a $\Kz$ and 
a $\Kzbar$ meson.
These contributions are expected to be small for $B$ mixing and 
are assumed to be negligible in this analysis.

We present a measurement of the \Bz\ lifetime \tauBz\ and the
oscillation frequency \Dm\ based on a sample of $\approx$14,000 
exclusively reconstructed \btodstlnu\ decays\footnote{Throughout this 
paper, charge conjugate modes are always implied.} 
selected from a 
sample of 23 million $B\overline B$ events recorded at the \FourS\ resonance
with the \babar\ detector at the Stanford Linear Accelerator Center,
in 1999-2000.
In this experiment, 9~GeV electrons and 3.1~GeV positrons, 
circulating in the PEP-II storage ring,
annihilate
to produce $B\overline B$ pairs moving along the $e^-$ beam
direction ($z$ axis) with a known Lorentz boost of $\beta\gamma = 0.55$,
which allows a measurement of the time between the two $B$ decays,
$\Dt$.

The proper decay-time difference $\Dt$ between two neutral $B$
mesons produced in a coherent $P$-wave state in an \FourS\ event
is governed by the following probabilities to observe an unmixed event,
 \begin{equation}
 P(\BzBzbar \rightarrow \BzBzbar) \propto 
 e^{-|\Dt|/\tauBz}(1 + \cos\Dm\Dt),
 \label{eq:trueP1}
  \end{equation} 
or a mixed event,
 \begin{equation}
 P(\BzBzbar \rightarrow \Bz\Bz\ {\rm or}\ \Bzbar\Bzbar) \propto 
 e^{-|\Dt|/\tauBz}(1 - \cos\Dm\Dt).
 \label{eq:trueP2}
  \end{equation} 
Therefore, if we measure $\Dt$ and identify the $b$-quark flavor
of both $B$ mesons at their time of decay, we can extract the \Bz\ lifetime
\tauBz\ and the mixing frequency \Dm.
In this analysis, one $B$ is reconstructed in the mode \btodstlnu, 
which has a measured  branching fraction of 
$(4.60 \pm 0.21)\%$~\cite{ref:PDG2002}.  
Although the neutrino cannot be detected, 
the requirement of a reconstructed $D^{*-}\rightarrow \overline D^0\pi^-$ 
decay and a
high-momentum lepton satisfying kinematic constraints consistent with the decay
\btodstlnu\ allows the isolation of a signal sample with 
(65 - 89)\% purity, depending on the $D^0$ decay mode
and whether the lepton candidate is an electron or a muon.
The charges of the
final-state particles identify the meson as a \Bz\ or a \Bzbar.
The remaining charged particles in the event, which originate from the other 
$B$ (referred to as \Btag), are used to identify, or ``tag", its 
flavor as a \Bz\ or a \Bzbar.
The time difference 
$\Dt = \tdstl - \ttag \approx \Dz/\beta\gamma c$
is determined from the separation $\Dz$ of the decay vertices 
for the \dstl\ candidate and the tagging $B$ along the boost direction.
The average separation is 250~$\mu$m.

The oscillation frequency \Dm\ and the average lifetime
of the neutral $B$ meson, \tauBz, are determined simultaneously with
an unbinned maximum-likelihood fit to the measured $\Dt$ distributions
of events that are classified as mixed and unmixed.
This is in contrast to published  measurements in which only 
\Dm\ is measured with \tauBz\ fixed to the world average, or
only \tauBz\ is measured.
There are several reasons to measure the lifetime and oscillation 
frequency simultaneously.
The statistical precision for both \tauBz\ and \Dm\ is comparable 
to the uncertainty on the  world average.  
Therefore, it is appropriate to measure both quantities rather than 
fixing the lifetime to the world average.
Since mixed and unmixed events have different \Dt\ distributions,
the mixing information for each event gives greater sensitivity to
the \Dt\ resolution function and a smaller statistical uncertainty on
\tauBz.
Also, since $B^+B^-$ events do not mix, 
we can use the \Dt\ distributions for mixed and unmixed
events to help discriminate between \BzBzbar\ signal events 
and $B^+B^-$ background events in the lifetime and mixing measurement.

There are three main experimental complications that affect the $\Dt$ 
distributions given in Eqs.~\ref{eq:trueP1} and \ref{eq:trueP2}.
First, the tagging algorithm, which classifies events into 
categories $c$ depending on the source of the available
tagging information, incorrectly identifies the flavor of \Btag\ with
a probability $w_c$ with a consequent reduction of the observed
amplitude for the mixing oscillation by a factor 
$(1-2w_c)$.
Second, the resolution for $\Dt$ is comparable to the 
lifetime and must be well understood.
The probability density functions (PDF's) for the unmixed ($+$) and mixed ($-$)
signal events can be expressed as the 
convolution of the underlying \Dttrue\ distribution for  
tagging category $c$, 
$$  {e^{-|\Dttrue|/\tauBz}\over 4\tauBz}
  [1\pm(1-2w_c)\cos\Dm\Dttrue],$$
with a resolution function 
${\cal R}(\Dtmeas - \Dttrue; \vec{q}_c)$
that depends on a set of parameters $\vec{q}_c$.
A final complication is that the sample of selected \btodstlnu\ candidates is
not pure signal.  

To characterize the backgrounds,
we select control samples of events enhanced
in each type of background
and determine the signal and background probabilities for each event in
the signal and background control samples
as described in Sec.~\ref{sec:evtsel}.
The measurement of \Dz and the determination of 
\Dt\ and \sigmaDt\ for each event is discussed in Sec.~\ref{sec:dectime}.
The $b$-quark tagging algorithm is described in Sec.~\ref{sec:tagging}.
In Sec.~\ref{sec:fitmodel}, we describe the unbinned maximum-likelihood fit.
The physics model and \Dt\ resolution function used to describe the 
measured \Dt\ distribution for signal are given in Sec.~\ref{sec:sigmodel}.
A combination of Monte Carlo simulation and data samples are used to determine
the parameterization of the PDF's to describe the \Dt\ distribution for each type
of background, as described in Sec.~\ref{sec:bkgndmodel}.
The likelihood is maximized in a simultaneous fit to the signal and 
background control samples to extract 
the \Bz\ lifetime \tauBz,
the mixing frequency \Dm, 
the mistag probabilities $w_c$, 
the signal \Dt\ resolution parameters $\vec{q}_c$,
the background \Dt\ model parameters,
and the fraction of $B^+\rightarrow\dstlnu X$ decays in the signal sample.
The results of the fit are given in Sec.~\ref{sec:results}.
Cross-checks are described in Sec.~\ref{sec:validations} and 
systematic uncertainties are summarized in Sec.~\ref{sec:systematics}.


\section{The \babar\ detector}

\label{sec:detector}

The \babar\ detector is described in detail elsewhere~\cite{ref:babar}.
The momenta of charged particles are measured with a
combination  of a 40-layer drift chamber (DCH) and a five-layer silicon vertex
tracker (SVT) in a 1.5-T solenoidal magnetic field.
A detector of internally-reflected Cherenkov radiation (DIRC) is used
for charged hadron identification.
Kaons are identified with a neural network based on the likelihood ratios 
calculated from d$E$/d$x$ measurements in the SVT and DCH, and from the observed
pattern of Cherenkov light in the DIRC.
A finely-segmented CsI(Tl) electromagnetic calorimeter (EMC) is used to detect photons
and neutral hadrons, and to identify electrons.
Electron candidates are required to have a 
ratio of EMC energy to track momentum, 
an EMC cluster shape,
DCH d$E$/d$x$, 
and DIRC Cherenkov angle all consistent with the electron hypothesis.
The instrumented flux return (IFR) contains resistive plate
chambers for muon and long-lived neutral hadron identification.
Muon candidates are required to have IFR hits
located along the extrapolated DCH track, 
an IFR penetration length,
and an energy deposit in the EMC consistent with
the muon hypothesis.


\section{Data samples}

\label{sec:samples}

The data used in this analysis were recorded with the \babar\
detector \cite{ref:babar} at the PEP-II storage ring \cite{ref:pepii} 
in the period October 1999 to December 2000.
The total integrated luminosity of the data set is 20.6~fb$^{-1}$ 
collected at the \FourS\ resonance and 2.6~fb$^{-1}$ collected 
about 40~MeV below
the \FourS\ (off-resonance data).
The corresponding number of produced $B\overline B$ pairs is 23 million.

Samples of Monte-Carlo simulated $B\overline B$  and $c\overline c$ events, generated
with a GEANT3~\cite{ref:GEANT3} detector simulation, are analyzed through the same 
analysis chain as the real data to check for biases in the extracted physics parameters
and are also used to develop  models for describing detector resolution effects.  
The values of the parameters used in these models are determined with real data. 
The equivalent luminosity
of this simulated data is approximately equal to that of the real data for
$B\overline B$ events and about half that of real data for $c\overline c$ events.
In addition, we generate signal Monte Carlo samples in which one neutral $B$ meson in
every event decays to \dstlnu, with \dsttodpi, and the other
neutral $B$ meson decays generically.   
The $D^0$ then decays to one of the four final states
reconstructed in this analysis (described in the next section).
The equivalent luminosity of the simulated signal samples is equal to approximately
2 to 8 times that of the real data, depending on the $D^0$ decay mode.


\section{Event selection and characterization}

\label{sec:evtsel}

We select events containing a fully-reconstructed 
\dstm\ and an identified oppositely-charged electron or muon. 
This $\dstl$ pair is then required to pass kinematic cuts that enhance the
contribution of semileptonic \btodstlnu\ decays.  
In  addition to the signal sample, we select several control samples that are used to
characterize the main sources of background.

We define the following classification of the sources of signal and background that we
expect to contribute to this sample.  
The nomenclature shown in italics will be used throughout this paper to define signal
and all possible types of background.
\begin{enumerate}
\item Events with a correctly reconstructed \dstm\ candidate:
 \begin{enumerate}
 \item Events that originate from $B\overline B$ events:
    \begin{enumerate}
    \item Events with a correctly identified lepton candidate:
     \begin{enumerate}
          \item {\it Signal} -- \btodstlnu(X) decays.
          \item {\it Uncorrelated-lepton background} -- 
           ($B\rightarrow D^{*-} X,\ {\rm other}\ {B}\rightarrow\ell^+ X$) or
           ($B\rightarrow D^{*-} X,\ X\rightarrow\ell^+ Y$)
          \item {\it Charged $B$ background} -- 
           $B^+\rightarrow D^{*-}\ell^+\nu_\ell X$.
     \end{enumerate}
    \item {\it Fake-lepton background} -- events with a misidentified lepton candidate.
    \end{enumerate}
    \item {\it Continuum background} -- $c\overline c \rightarrow D^{*-} X$.
 \end{enumerate}
 \item {\it Combinatoric background} -- events with a misreconstructed 
 \dstm\ candidate.
\end{enumerate}

Lepton candidates are defined as charged tracks with momentum in the \FourS\
rest frame greater than 1.2~GeV.
For the \dste\ samples, the electron candidate passes selection
criteria with a corresponding electron identification efficiency of about 90\% and
hadron misidentification less than 0.2\%.
For the \dstmu\ samples, the muon candidate passes selection
criteria with a corresponding muon identification efficiency of about 70\% and
hadron misidentification between 2\% and 3\%.
For the fake-lepton control sample, \dstl\ candidates are accepted if the lepton {\em fails}
both electron and muon selection criteria looser than those required for lepton
candidates.

\dst\ candidates are selected in the decay mode \dsttodpi. 
The $D^0$ candidate is reconstructed in the modes
$K^-\pi^+$,
$K^-\pi^+\pi^-\pi^+$,
$K^-\pi^+\pi^0$ and $\ks\pi^-\pi^+$. 
The daughters of the $D^0$ decay are selected according to the following definitions.
$\pi^0$ candidates are reconstructed from two photons with invariant mass within 
15.75~MeV of the $\pi^0$ mass. 
The mass of the photon pair is constrained to the $\pi^0$ mass and 
the photon pair is kept as a $\pi^0$ candidate if the \chisq\ probability of the fit
is greater than 1\%.
\ks\ candidates are reconstructed from a pair of charged particles with 
invariant mass within 15~MeV of the \ks\ mass. 
The pair of tracks is retained as a \ks\ candidate if the \chisq\ probability that the
two tracks form a common vertex is greater than 1\%.
Charged kaon candidates satisfy loose kaon criteria for the
$K^-\pi^+$ mode and tighter criteria for the $K^-\pi^+\pi^-\pi^+$ and $K^-\pi^+\pi^0$
modes. 
For the $K^-\pi^+\pi^0$ and $\ks\pi^-\pi^+$ modes, a likelihood is
calculated as the square of the decay amplitude in the Dalitz plot for the three-body
candidate, based on measured amplitudes and phases \cite{ref:E687}.
The candidate is retained if the likelihood is greater than 10\% of its maximum value
across the Dalitz plot.

$D^0$ candidates have measured invariant mass within 17~MeV of the $D^0$ mass for the 
$K^-\pi^+$, $K^-\pi^+\pi^-\pi^+$, and $\ks\pi^-\pi^+$ modes, and within 34~MeV for the
$K^-\pi^+\pi^0$ mode.
The invariant mass of the daughters is constrained to the $D^0$ mass and the tracks
are constrained to a common vertex in a simultaneous fit.  
The $D^0$ candidate is retained if the \chisq\ of the fit is greater than 0.1\%.

The low-momentum pion candidates for the \dsttodpi\ decay are selected
with total momentum less
than  450~MeV in the \FourS\ rest frame
and momentum transverse to the beamline greater than 50~MeV.
The momentum of the \dst\ candidate in the \FourS\ rest frame 
is between 0.5 and 2.5~GeV.
The \dstl\ candidate satisfies $|\cos\theta^*_{\rm thrust}| < 0.85$, where
$\theta^*_{\rm thrust}$ is the angle between the thrust axis of the \dstl\ candidate
and the thrust axis of the remaining charged and neutral particles in the event.
\dstl\ candidates are retained if the \chisq\
probability that the daughter tracks form a common vertex is greater than 1\%
and
$\massdiff$ is less than 165~MeV,
where $m(D^*)$ is the candidate $\Dzbar\pi^-$ mass calculated with 
the candidate $\Dzbar$ mass constrained to the true $D^0$ mass, 
$m(D^0)$.

We define two angular quantities for each \dstl\ candidate to classify 
them into a sample 
enriched in \btodstlnu\ signal events, in which the \dst\ and lepton
candidates are on opposite sides of the event, and 
a sample enriched in {\it uncorrelated-lepton
background} events, in which the \dst\ and lepton candidates are on the same 
side of
the event. 
The first angle is $\theta_{\dst,\ell}$, the angle between the \dst\ and
lepton candidates in the \FourS\ rest frame.  
The second angle is \thby, the angle between the direction of the
\Bz\ and the vector sum of the \dst\ and lepton candidate momenta, 
calculated in the \FourS\ rest
frame. 
Since we do not know the direction of the \Bz, we calculate the cosine of \thby\ from
the following equation, in which we assume that the only $B$ decay particle missed in
the reconstruction is a massless neutrino:
 \begin{equation}
 \costhby \equiv {-(m^2_{\Bz} + m^2_{\dst\ell} - 2 E_B E_{\dst\ell}) \over 
              2 |\vec{p}_B| |\vec{p}_{\dst\ell}|}.
 \label{eq:costhby}
 \end{equation} 
All quantities in Eq.~\ref{eq:costhby} are defined in the \FourS\ rest frame.
The energy and the magnitude of the momentum of the $B$ are 
calculated from the 
$e^+e^-$ center-of-mass energy and the \Bz\ mass.
For true \btodstlnu\ events, \costhby\ lies in the physical
region $[-1, +1]$, except for detector resolution effects.
Backgrounds lie inside and outside the range $[-1, +1]$.
We also calculate the same angle with the lepton momentum direction 
reflected through the origin in the \FourS\ rest frame: \thbyfl.

A sample enhanced in \btodstlnu\ signal events 
(called the {\it opposite-side} sample)
is composed of \dstl\ candidates with $\cos\theta_{\dst\ell}<0$ and 
$|\costhby|<1.1$.
Samples are defined for lepton candidates that satisfy the criteria for an electron, a
muon and a fake-lepton.
The first two samples are the signal samples, and the latter is the
{\it fake-lepton} control sample.

An additional background control sample, representative of the 
uncorrelated-lepton background and called the {\it same-side} sample, is composed of 
\dstl\ candidates satisfying $\cos\theta_{\dst\ell}\geq 0$ and $|\costhbyfl|<1.1$. 
We use \costhbyfl\ rather than \costhby\ because, in 
Monte Carlo simulation,
the distribution of \costhbyfl\ in this
control sample is similar to the distribution of \costhby\ for
uncorrelated-lepton background in the signal sample, whereas the distribution of
\costhby\ in the background control sample is systematically different.

\dstl\ candidates are retained if a fit of the lepton, $\pi^-$, and $D^0$ candidates to
a common vertex converges.
In addition, several criteria that depend on the charged tracks in the rest of the
event, as well as the \dstl\ candidate, are applied.  
Events are retained if at least two tracks are used to determine the decay point of the other
$B$, 
the fit that determines the distance \Dz\ between the two $B$
decays along the beamline converges, the time between decays  (\Dt) calculated from
\Dz is less than 18~ps, and the calculated error on \Dt (\sigmaDt) is less than
1.8~ps.

Approximately $68,000$ candidates pass the above selection criteria. 
These candidates are distributed over two signal samples 
and ten background control samples
defined by the following characteristics:
whether the data was recorded on or off the \FourS\ resonance (two choices);
whether the candidate lepton is {\it same-side} or {\it opposite-side} to the \dst\
candidate (two choices); and
whether the lepton candidate passes the criteria for an electron, a muon, or a fake
lepton (three choices).

\begin{figure}[!htb]
\begin{center}
\includegraphics[width=3in]{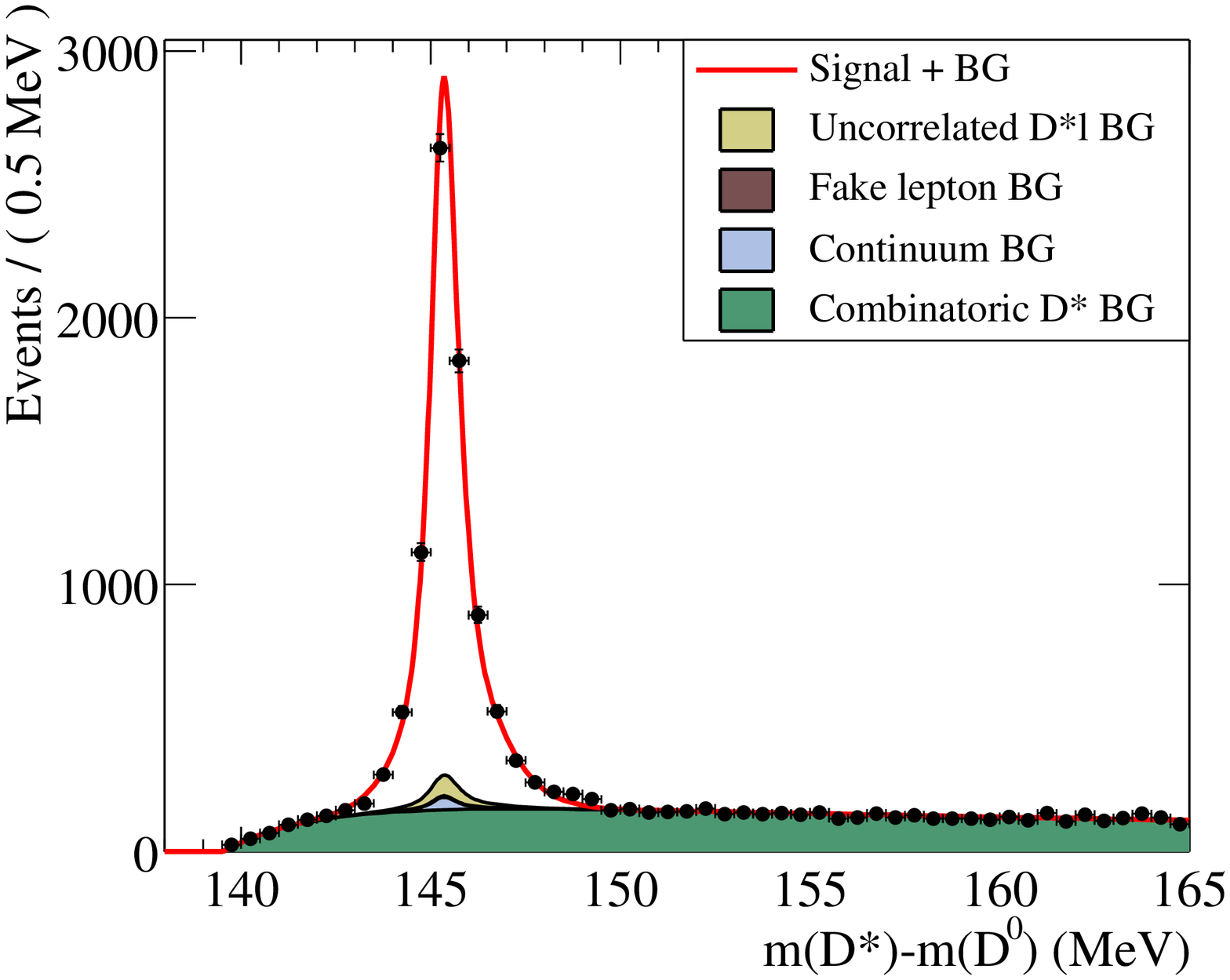}
\includegraphics[width=3in]{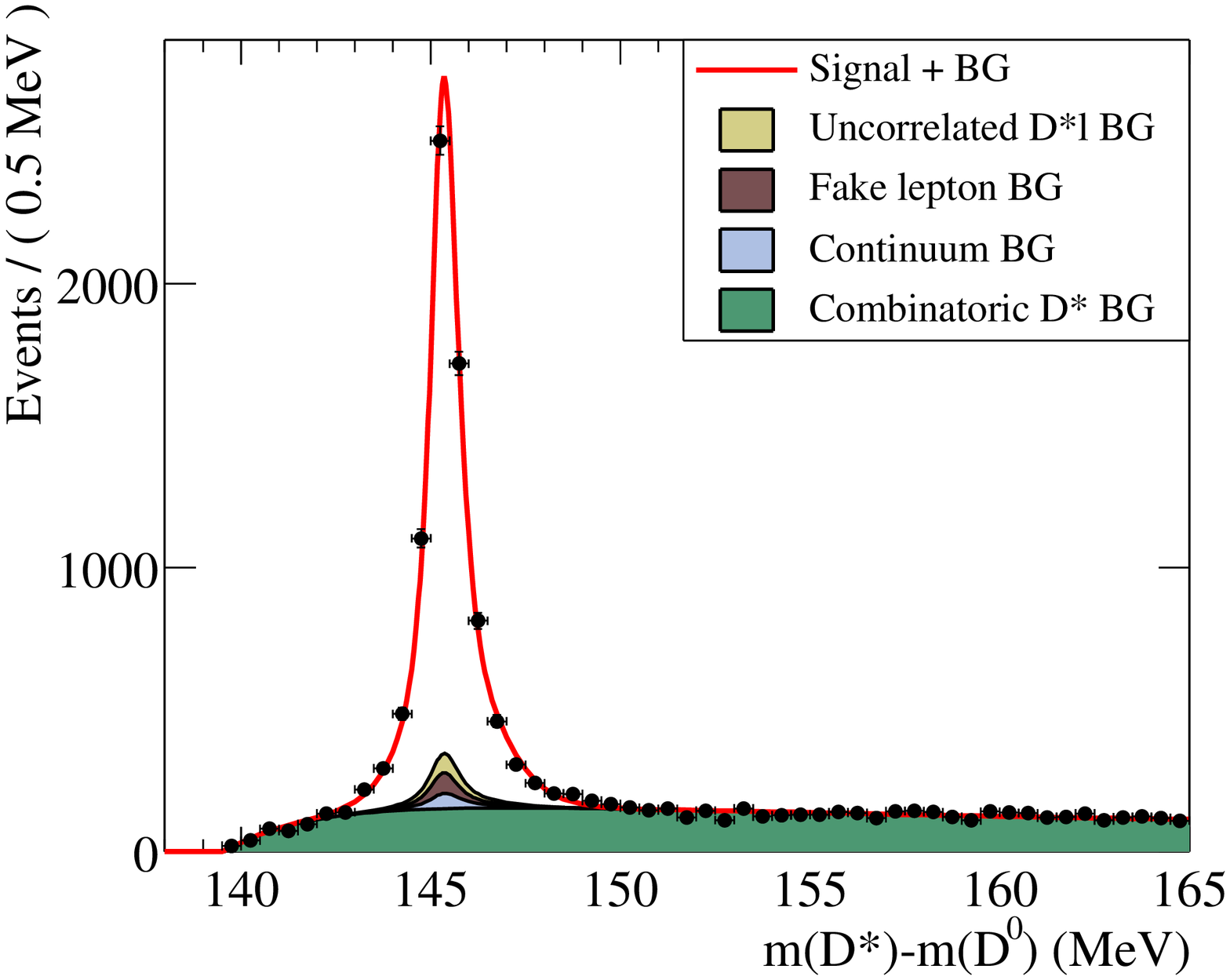}
\caption{\massdiff\ distribution for events passing all selection criteria for
\btodstlnu, with an electron (left) or muon (right) candidate.  
The points correspond to the data.
The curve is the result of a simultaneous unbinned maximum likelihood fit to
this sample of events and a number of background control samples.
The shaded distributions correspond to the four types of background 
described in the text.}
\label{fig:massdifsig}
\end{center}
\end{figure}

The combinatoric background due to events with a misreconstructed \dst\
candidate can be distinguished from events with a real \dst\ in a plot 
of the mass difference \massdiff.
The \massdiff\ distributions for the samples of signal events 
(opposite-side \dste\ and \dstmu\ candidates in on-resonance data) 
are shown as data points in
Fig.~\ref{fig:massdifsig} for electron candidates (left) 
and muon candidates (right).

Each of the 12 samples described above 
are further divided into 30 subsamples
according to the following characteristics that may affect the \massdiff\ or \Dt\
distributions.
\begin{enumerate}
\item The $\pi^-$ from the \dst\ decay reconstructed in the SVT only, or in the SVT
and DCH (two choices):  The \massdiff\ resolution is worse when the $\pi^-$ is
reconstructed only in the SVT.
\item The $D^0$ candidate reconstructed in the mode $K^-\pi^+$ or $K^-\pi^+\pi^0$ or
($K^-\pi^+\pi^-\pi^+$ or $\ks\pi^-\pi^+$) (three choices): The level of contamination
from combinatoric background and  the \massdiff\ resolution may depend on the $D^0$ decay
mode. 
\item The $b$-tagging information used for the other $B$ (five choices; see Sec.~\ref{sec:tagging}):
The level of contamination from each type of background and the \Dt\ resolution parameters may
depend on the tagging information.
\end{enumerate} 
This allows subdivision into 360 samples.
In the unbinned maximum likelihood fits to the \massdiff\ and (\Dt, \sigmaDt)
distributions, individual fit parameters are shared among different sets of subsamples
based on physics motivation and observations in the data.

We do a simultaneous fit to the \massdiff\ distributions for all 360 subsamples.
The peak due to real \dst\ candidates is modeled by the sum of two Gaussian
distributions; 
the mean and variance of both the Gaussian distributions, as well as the relative
normalization of the two Gaussians, are free parameters in the fit. 
We model the shape
of the combinatoric background with the function
 \begin{equation}
 {1\over N} \left[ 1-\exp\left(-{\dm - \dm_0\over c_1}\right)\right] 
  \left({\dm\over \dm_0}\right)^{c_2},
 \label{eq:combback}
 \end{equation} 
where $\dm \equiv \massdiff$, $N$ is a normalization constant, $\dm_0$ is a
kinematic threshold equal to the mass of the $\pi^-$, and $c_1$ and $c_2$ are free
parameters in the fit.  
An initial fit is done to determine the shape parameters describing the peak and
combinatoric background. 
Separate values of the five parameters describing the shape of the peak are used
for the six subsamples defined by  whether the $\pi^-$ candidate is tracked in the SVT
only or in the SVT and DCH, and the three types of $D^0$ decay modes.
Each of these six groups that share peak parameters is further subdivided
into twelve subgroups that each share a common set of the two combinatoric
background shape parameters.  
Ten of these twelve subgroups are defined by the five tagging categories for the large signal samples and for the
fake-lepton control samples, in on-resonance data.
The other two subgroups are defined as
same-side,
on-resonance samples and 
all off-resonance samples.

Once the peak and combinatoric background shape parameters have been determined, 
we fix the shape parameters and determine the peak and combinatoric background
yields in each of the 360 subsamples with an unbinned extended maximum-likelihood fit.
The total peak yields in the signal sample and each background control sample are then used
to determine the amount of true signal and each type of peaking background in the
\massdiff\ peak of each sample as follows.
\begin{enumerate}
\item
{\em Fake-lepton background --} \ \ 
Particle identification and misidentification efficiencies 
for the electron, muon, and fake-lepton selection criteria
are measured in data as a
function of laboratory momentum, polar angle, and azimuthal angle, for 
true electrons, muons, pions, kaons, and protons.  
$\Bz\Bzbar$ and $B^+B^-$ Monte Carlo simulations are used to determine the 
measured laboratory momentum,
polar angle, and azimuthal angle distributions for true 
electrons, muons, pions, kaons
and protons that pass all  selection criteria for \dstl\ candidates,
except the lepton (or fake-lepton) identification criteria. 
These distributions are combined with the measured particle identification
efficiencies and misidentification probabilities
to determine the momentum- and
angle-weighted probabilities for a true lepton or true hadron to pass the criteria for
a lepton or a fake lepton in each of the \dstl\ signal and background control samples.
We then use these efficiencies and misidentification probabilities, 
and the observed
number of lepton and fake-lepton candidates in data, to determine the number of
true leptons and fake leptons (hadrons) in each control sample.
\item
{\em Uncorrelated-lepton background  --}\ \ 
To determine the number of uncorrelated-lepton events in each sample, we use the
relative efficiencies from Monte Carlo simulation for signal and uncorrelated-lepton
events to pass the criteria for same-side and opposite-side samples.
\item
{\em Continuum background  --}\ \ 
We use the peak yields in off-resonance data, scaled by the 
relative integrated luminosity for
on- and off-resonance data, to determine the continuum-background yields in
on-resonance data.
\end{enumerate}

The peak yields and continuum, fake-lepton, and uncorrelated-lepton fractions of
the peak yield, as well as the combinatoric fraction of all events in a \massdiff\
signal window, are shown in Table~\ref{tab:yields} 
for the signal and background control samples in on-resonance data.  
The peak yields include the peaking backgrounds. 
The signal window is defined as (143 - 148)~MeV for the calculation of combinatoric
background fractions.

\begin{table}[!htb]
\caption{Peak yields and continuum, fake-lepton, and uncorrelated-lepton fractions of
the peak yield, and the combinatoric fraction of total events in a \massdiff\
signal window for the signal and background control samples 
in on-resonance data.  Peak yields
include the peaking backgrounds.  The signal window for combinatoric background
fractions is defined as (143 - 148)~MeV.
OS and SS refer to opposite-side and same-side samples;
$e$, $\mu$, and f indicate the type of lepton candidate: electron, muon 
or fake-lepton.}
\begin{center}
\begin{tabular}{l|l|rrrr|r}
\hline\hline
\multicolumn{2}{c|}{Category}
& Peak Yield & $f_{\mathrm{cont}}(\%)$ & $f_{\mathrm{fake}}(\%)$
& $f_{\mathrm{uncorr}}(\%)$ & $f_{\mathrm{comb}}(\%)$ \\
\hline
OS & $e$ & $ 7008\pm 91$ & $ 1.53\pm 0.42$ & $ 0.1678\pm 0.0042$ & $ 3.14\pm 0.39$ & $ 17.89\pm 0.24$ \\
& $\mu$ & $ 6569\pm 88$ & $ 2.27\pm 0.57$ & $ 2.669\pm 0.067$ & $ 2.85\pm 0.48$ & $ 18.36\pm 0.25$ \\
& f & $ 8770\pm 108$ & $ 12.8\pm 1.3$ & $ 72.4\pm 1.8$ & $ 0.7\pm 1.6$ & $ 31.40\pm 0.24$ \\
\cline{1-7}
SS & $e$ & $ 306\pm 21$ & $ 0.000\pm 0.006$ & $ 0.533\pm 0.039$ & $ 56.9\pm 7.0$ & $ 34.0\pm 1.3$ \\
& $\mu$ & $ 299\pm 20$ & $ 5.1\pm 3.6$ & $ 8.89\pm 0.64$ & $ 48.9\pm 8.0$ & $ 34.4\pm 1.3$ \\
& f & $ 1350\pm 45$ & $ 20.4\pm 4.1$ & $ 74.4\pm 5.4$ & $ 3.6\pm 7.8$ & $ 42.59\pm 0.61$ \\
\hline
\hline
\end{tabular}
\end{center}
\label{tab:yields}
\end{table}

Finally, we use the calculated fractions and fitted shapes of the background sources
in each control sample to
estimate the probability of each candidate to be due to signal or each type of
background (combinatoric, continuum, fake-lepton, or uncorrelated-lepton)
when we fit the (\Dt, \sigmaDt) distribution to determine the lifetime and mixing
parameters.   
We take advantage of the fact that charged and neutral $B$ decays have
different decay-time distributions (because the charged $B$ does not mix) to 
determine the fraction of charged $B$ background events in the fit to 
(\Dt, \sigmaDt).


\section{Decay-time measurement}

\label{sec:dectime}

The decay-time difference \Dt\ between $B$ decays is determined from the measured
separation $\Dz = \zdstl - \ztag$ along the $z$ axis 
between the \dstl\ vertex position
(\zdstl) and the flavor-tagging decay \btag\ vertex position (\ztag). 
This measured \Dz\ is converted into \Dt\ with the use of the known \FourS\ boost,
determined for each run.
Since we cannot reconstruct the direction of the $B$ meson for each event, 
we use the excellent approximation $\Dt \approx \Dz/(\beta\gamma c)$.

The momentum and position vectors of the \dz, $\pi^-$, and lepton candidates, and the 
average position of the $e^+e^-$ interaction point (called the beam spot) in the plane
transverse to the beam  are used in a constrained fit to determine the
position of the \dstl\ vertex. 
The beam-spot constraint is of order 100~$\mu$m in the horizontal
direction and  30~$\mu$m in the vertical direction, corresponding to the RMS size of
the beam in the horizontal direction and the approximate transverse flight path of the
$B$ in the vertical direction.
The beam-spot constraint improves the resolution on \zdstl\ by about 30\%.
The RMS spread on the difference between the measured and true position
of the \dstl\ vertex is about 80~$\mu$m (0.5~ps).

We determine the position of the \btag\ vertex from all charged tracks
in the  event except the daughters of the \dstl\ candidate, using \ks\ and
$\Lambda$ candidates in place of their daughter tracks, and excluding tracks that are
consistent with being due to photon conversions.  
The same beam-spot constraint applied to the \bdstl\ vertex is also applied to the 
\btag\ vertex.
To reduce the influence of charm decay products, which bias the determination of the
vertex position,  tracks with a large contribution to the \chisq\ of the vertex fit are
iteratively removed until those remaining have a reasonable fit probability
or only one track remains.  
The RMS spread on the difference between the measured and true position
of the \btag\ vertex is about 160~$\mu$m (1.0~ps).
Therefore, the \Dt\ resolution is dominated by the $z$ resolution of
the tag vertex position. 
Events are retained if the fit converges, 
at least two tracks contribute to the determination of the tag vertex fit,  
the time between decays  (\Dt) calculated from \Dz\ is less than 18~ps,
and the calculated error on \Dt (\sigmaDt) is less than 1.8~ps.

We calculate the uncertainty on \Dz\ due to uncertainties on the track parameters from the SVT and DCH hit resolution and multiple scattering, our knowledge of the beam-spot
size, and the average $B$ flight length in the vertical direction.
The calculated uncertainty does not account for errors in 
pattern recognition in
tracking, 
errors in associating tracks with the $B$ vertex, or
the effects of misalignment within and between the tracking devices.
The calculated uncertainties will also be incorrect if 
our assumptions for the amount
of material in the tracking detectors or the beam-spot size or 
position are inaccurate.
We use parameters in the \Dt\ resolution model, measured with data, to account for
uncertainties and biases introduced by these effects.


\section{Flavor tagging}

\label{sec:tagging}

All charged tracks in the event, except the daughter tracks of the \dstl\ candidate,
are used to determine whether the \btag\ decayed as a \Bz\ or a \Bzbar. 
This is called flavor tagging.
We use five different types of flavor tag, or tagging categories, in this analysis.
The first two tagging categories rely on the presence of a prompt lepton, or one or
more charged kaons, in the event. 
The other three categories exploit a variety of inputs with a neural-network algorithm.
The tagging algorithms are described briefly in this section; 
see Ref.~\cite{ref:PRD} for more details.

Events are assigned a \lepton\ tag if they contain an identified lepton with momentum
in the \FourS\ rest frame greater than 1.0 or 1.1 GeV for electrons and muons,
respectively, thereby selecting mostly primary leptons from the decay of the
$b$ quark.
If the sum of charges of all identified kaons is nonzero, the event is assigned a
\kaon\ tag.
The final three tagging categories 
are based on the output of a neural network that uses as inputs
the momentum and charge of the track with the maximum center-of-mass momentum,
the number of charged tracks with significant impact parameters with respect 
to the interaction point, 
and the outputs of three other neural networks,
trained to identify primary leptons, kaons, and soft pions. 
Depending on the output of the main neural network, 
events are assigned to an \ntone\ (most
certain), \nttwo, or \ntthree\ (least certain) tagging category. 
About 30\% of events are in the \ntthree\ category, which has a mistag rate close to
50\%.  
Therefore, these events do not carry much sensitivity to the mixing 
frequency, but they
increase the sensitivity to the \Bz\ lifetime.

Tagging categories are mutually exclusive due to the hierarchical use of the tags.
Events with a \lepton\ tag and no conflicting \kaon\ tag are assigned to the
\lepton\ category.
If no \lepton\ tag exists, but the event has a \kaon\ tag, it is assigned to the 
\kaon\ category.
Otherwise events are assigned to corresponding neural network
categories.


\section{Fit method}

\label{sec:fitmodel}

We perform an unbinned fit simultaneously to events in each of the 12
signal and control samples (indexed by $s$) that are further
subdivided into 30 subsamples (indexed by $c$)
using a likelihood
\begin{equation}
{\cal L} = \prod_{s=1}^{12}\, \prod_{c=1}^{30}\,\prod_{k=1}^{N(s,c)}\,
P_{s,c}(\vec{x}_{k}\,;\vec{p}) \; ,
\label{eq:likelihood}
\end{equation}
where $k$ indexes the $N(s,c)$ events $\vec{x}_{k}$ in each of the 360
subsamples. 
The probability $P_{s,c}(\vec{x}_{k}\,;\vec{p})$ of observing an event 
$\vec{x}_k = (\delta m,\Dt,\sigmaDt,g)$ 
is calculated as a function of the parameters 
$\vec{p} = (f_{s,c}^{\,\mathrm{comb}}, {\vec{p}}_{s,c}^{\,\mathrm{comb}},
{\vec{p}}_{c}^{\,\mathrm{peak}}, {\vec{q}}_{s,c}^{\,\mathrm{comb}},
f_{s,c,1}^{\,\mathrm{pkg}}, f_{s,c,2}^{\,\mathrm{pkg}}, f_{s,c,3}^{\,\mathrm{pkg}},
{\vec{q}}_{s,c,1}^{\,\mathrm{pkg}}, {\vec{q}}_{s,c,2}^{\,\mathrm{pkg}},
{\vec{q}}_{s,c,3}^{\,\mathrm{pkg}}, {\vec{q}}_{c}^{\,\mathrm{sig}})$ as
\begin{multline}
P_{s,c}(\delta m,\Dt,\sigmaDt,g\,;\vec{p}) =\\
f_{s,c}^{\,\mathrm{comb}}\cdot {\cal F}^{\,\mathrm{comb}}(\delta m\,;\vec{p}_{s,c}^{\,\mathrm{comb}})\cdot
{\cal G}^{\,\mathrm{comb}}(\Dt,\sigmaDt,g\,;\vec{q}_{s,c}^{\,\mathrm{comb}}) +
\left(1 - f_{s,c}^{\,\mathrm{comb}}\right)\cdot {\cal F}^{\,\mathrm{peak}}(\delta
m\,;\vec{p}_{c}^{\,\mathrm{peak}})\cdot \\
\left[ \sum_{j=1}^3\, f_{s,c,j}^{\,\mathrm{pkg}}\cdot
{\cal G}_{j}^{\,\mathrm{pkg}}(\Dt,\sigmaDt,g\,;\vec{q}_{s,c,j}^{\,\mathrm{pkg}}) +
\left(1 - \sum_{j=1}^3\,
f_{s,c,j}^{\,\mathrm{pkg}}\right)\cdot
{\cal G}^{\mathrm{sig}}(\Dt,\sigmaDt,g\,;\vec{q}_{c}^{\,\mathrm{sig}})
\right] \;,
\label{eq:master}
\end{multline}
where $j$ indexes the three sources of peaking background and
\dm\ is the mass difference \massdiff\ defined earlier.
The index $g$ is $+1$ ($-1$) for unmixed (mixed) events.
By allowing different effective mistag
rates for apparently mixed or unmixed events
in the background functions ${\cal G}^{\,\mathrm{comb}}$ and ${\cal G}^{\,\mathrm{pkg}}$, 
we accomodate the different levels of backgrounds observed in mixed and unmixed samples.
Functions labeled with ${\cal F}$ describe the probability of
observing a particular value of $\delta m$ while functions labeled with
${\cal G}$ give probabilities for values of $\Dt$ and $\sigmaDt$
in category $g$.
Parameters labeled with $f$
describe the relative contributions of different types of
events. Parameters labeled with $\vec{p}$ describe the shape of a
$\delta m$ distribution, and those labeled with $\vec{q}$ describe a
(\Dt, \sigmaDt) shape.

Note that we make explicit assumptions that the $\delta m$ peak
shape, parameterized by $\vec{p}_{c}^{\,\mathrm{peak}}$, and the signal
(\Dt, \sigmaDt) shape, parameterized by $\vec{q}_{c}^{\,\mathrm{sig}}$, depend only
on the subsample index $c$. The first of these assumptions is
supported by data, and simplifies the analysis of peaking background
contributions. The second assumption reflects our expectation that the
\Dt\ distribution of signal events does not depend on whether they are
selected in the signal sample or appear as a background in a control sample.

The ultimate aim of the fit is to obtain the \Bz lifetime and mixing
frequency, which by construction are common to all sets of signal
parameters $\vec{q}_{c}^{\,\mathrm{sig}}$. Most of the statistical power for
determining these parameters comes from the signal sample, although
the fake and uncorrelated background control samples also contribute
due to their signal content (see Table~\ref{tab:yields}).

We bootstrap the full fit with a sequence of initial fits using reduced
likelihood functions to a partial set of samples, to determine the appropriate 
parameterization of the signal resolution function and the background \Dt\ models, 
and to determine starting values for each parameter in the full fit.

\begin{enumerate}
\item
We first find a model that describes the \Dt\ distribution for each type of 
event: signal, combinatoric background, and the 
three types of backgrounds that 
peak in the \massdiff\ distribution.
To establish a model, we use Monte Carlo samples that have been selected
to correspond to only one type of signal or background event
based on Monte Carlo truth information.
These samples are used to determine the \Dt\ model and 
the categories of events ({\it e.g.,}
tagging category, fake or real lepton) that 
can share each of the parameters in  the model. 
Any subset of parameters can be shared among any subset of the 360 subsamples.
We choose parameterizations and sharing of parameters that minimizes the number of
different  parameters while still providing an
adequate description of the \Dt\ distributions.
\item
We then find the starting values for  the background 
parameters by fitting to each of the background-enhanced control samples
in data, using the model (and sharing of parameters) determined 
in the previous step.
Since these background control samples are not pure, we start with the purest
control sample (combinatoric background events from the \massdiff\ 
sideband) and move on to less pure control samples, always using the models
established from earlier steps to describe the \Dt\ distribution of 
the contamination from other backgrounds.
\end{enumerate}

The result of the above two  steps is a \Dt model for each type of event and
a set of starting values for all parameters in the fit.
When we do the final fit, we fit all signal and control samples
simultaneously ($\approx$68k events), leaving all parameters
free in the fit (72 free parameters).
The physics parameters \tauBz\ and \Dm\ were kept hidden 
until all analysis details and the systematic errors 
were finalized, to eliminate experimenter's bias.
However, statistical errors on the parameters and changes 
in the physics parameters due to changes in the analysis were not hidden.


\section{Signal \Dt\ model}

\label{sec:sigmodel}

For signal events in 
a given tagging category $c$, the probability density function (PDF)
for \Dt\ 
consists of a physics model convolved with a \Dt\ resolution function:
$$
{\cal G}^{\mathrm{sig}}(\Dt,\sigmaDt,g\,;\vec{q}_c^{\,\mathrm{sig}}) 
= \left\{ \frac{1}{4\tauBz}
e^{-|\Dttrue|/\tauBz}(1+ g (1-2 \omega_c)\cos(\Dm\Dttrue))\right\} \otimes
{\cal R}(\dDt,\sigmaDt; \vec{q_c}) \, ,
$$
where ${\cal R}$ is a resolution function, which can be different for
different event categories, 
$g$ is +1 ($-1$) for unmixed (mixed) events,
and \dDt\ is the residual $\Dt - \Dttrue$.
The physics model shown in the above equation has seven parameters:
\Dm, \tauBz, and mistag fractions $\omega_c$ for each of the five tagging 
categories. 
To account for an observed correlation between the mistag rate and \sigmaDt in
the kaon category (described in Sec.~\ref{sec:verttag}), we
allow the mistag rate in the kaon category to vary as a linear function of
\sigmaDt:
\begin{equation}
\omega_\mathrm{kaon}= m_\mathrm{kaon} \cdot\sigmaDt +
\omega\mathrm{_{kaon}^{offset}}\, .
\label{eq:mistag}
\end{equation}
 In addition, we allow the
mistag fractions for \Bz tags and \Bzbar tags to be different. 
We define $\Delta \omega= \omega_{\Bz}- \omega_{\Bzbar}$ and 
$\omega=(\omega_{\Bz}+ \omega_{\Bzbar})/2$, so that 
$$
\omega_{\Bz / \Bzbar} = \omega \pm \frac{1}{2}\Delta \omega \, .
$$
There are 13 free parameters in the complete physics model for all tagging 
categories.

For the \Dt\ resolution model,
we use the sum of a single Gaussian distribution and the same
Gaussian convolved with a one-sided exponential to describe
the core part of the resolution function, 
plus a single Gaussian distribution to describe the contribution of 
``outliers'' --- events in which the reconstruction error \dDt\ is not 
described by the calculated uncertainty \sigmaDt:
\begin{equation}
\begin{split}
& {\cal
R}_\mathrm{GExp+G}(\dDt,\sigmaDt;s,\kappa,f,b^\mathrm{out},s^\mathrm{out},f^\mathrm{out})
\\ & = f\cdot G(\dDt; 0,s\sigmaDt) + (1-f-f^\mathrm{out})\cdot
G(u-\dDt; 0,s\sigmaDt)\otimes E(u;\kappa\sigmaDt) 
\\ & + f^\mathrm{out}\cdot
G(\dDt; b^\mathrm{out},s^\mathrm{out})\; ,
\label{eq:GExp}
\end{split}
\end{equation}
where $u$ is an integration variable in the convolution $G\otimes E$.
The functions $G$ and $E$ are defined by 
$$G(x;x_0,\sigma) \equiv {1\over\sqrt{2\pi}\sigma}
  \exp\left(-(x-x_0)^2/(2\sigma)^2\right)$$
and
$$E(x;a) \equiv 
 \left\{ \begin{array}{ll}
 {1\over a}\exp{(x/a)} & {\rm if}\ x\leq0, \\
 0           & {\rm if}\ x>0.\end{array}
 \right.$$

Since the outlier contribution is not expected to be described by the calculated 
error on each event,
the last Gaussian term in Eq.~\ref{eq:GExp} does not depend on \sigmaDt.
However, in the terms that describe the core of the resolution function 
(the first two terms on the right-hand side of Eq.~\ref{eq:GExp}), 
the Gaussian width $s$ and the effective decay constant $\kappa$ are scaled by
\sigmaDt. 
The scale factor $s$ is introduced to accommodate an overall underestimate
($s>1$) or overestimate ($s<1$) of the errors for all events.
The decay constant $\kappa$ is introduced to account for residual charm decay products 
included in the \btag\ vertex; $\kappa$ is scaled by \sigmaDt\ to account for
a correlation observed in Monte Carlo simulation  between the mean of the \dDt\ distribution
and the measurement error \sigmaDt.  
This correlation is due to the fact that, in $B$ decays, the vertex 
error ellipse for the $D$ decay products is oriented with its major axis along the $D$
flight direction, leading to a correlation between the $D$ flight direction and the
calculated uncertainty on the vertex position in $z$ for the \btag\ candidate.
In addition, the flight length of the $D$ in the $z$ direction is correlated with its
flight direction.
Therefore, the bias in the measured \btag\ position due to including $D$ decay
products is correlated with the $D$ flight direction.
Taking into account these two correlations, we conclude that $D$ mesons that have a
flight direction perpendicular to the $z$ axis in the laboratory frame will have the
best $z$ resolution and will introduce the least bias in a measurement of the $z$
position of the \btag\ vertex, while $D$ mesons that travel forward in the laboratory
will have poorer $z$ resolution and will introduce a larger bias in the measurement of
the \btag\ vertex.

The mean and RMS spread of \Dt\ residual distributions in Monte Carlo simulation
vary significantly among tagging categories.
We find that we can account for these differences by allowing the core 
Gaussian fraction $f$ to be different for each tagging category.
In addition, we find that the correlations among the three parameters 
describing the outlier Gaussian ($b^\mathrm{out}$, $s^\mathrm{out}$, $f^\mathrm{out}$)
are large and that the outlier parameters are highly correlated with other
resolution parameters. 
Therefore, we fix the outlier bias $b^\mathrm{out}$ and scale factor 
$s^\mathrm{out}$, and vary them over a wide range to evaluate 
the systematic uncertainty on the physics parameters due to fixing these parameters
(see Sec.~\ref{sec:systematics}).
The resolution model then has 8 free parameters: $s$, $\kappa$, $f^\mathrm{out}$,
and five fractions $f_c$ (one for each tagging category $c$).

As a cross-check, we also use a resolution function that is the sum of
a narrow and a wide  Gaussian distribution, and a third Gaussian to describe 
outliers:
\begin{align*}
& {\cal R}_\mathrm{G+G+G}(\delta\Dt,\sigmaDt;b,s,f,b^w,s^w,b^\mathrm{out},s^\mathrm{out},f^\mathrm{out}) \\
& =  f\cdot G(\dDt; b\sigmaDt,s\sigmaDt)
+ (1-f-f^\mathrm{out})\cdot G(\dDt; b^w\sigmaDt,s^w\sigmaDt) +
f^\mathrm{out}\cdot G(\dDt; b^\mathrm{out},s^\mathrm{out})\; .\\
\end{align*}


\subsection{Vertex-tagging correlations}

\label{sec:verttag}

A correlation of about 0.12~ps$^{-1}$ is observed between the mistag rate and the 
\Dt\ resolution for \kaon\ tags.
This effect is modeled in the resolution function for signal as a linear dependence
of the mistag rate on \sigmaDt, as shown in Eq.~\ref{eq:mistag}.
In this section, we describe the source of this correlation.

We find that both the mistag rate for \kaon\ tags and the calculated error on \Dt\ 
depend inversely on $\sqrt{\Sigma p_t^2}$, where $p_t$ is the transverse momentum with
respect to the $z$ axis of tracks from the \btag\ decay.
Correcting for this dependence of the mistag rate removes most of the correlation
between the mistag rate and \sigmaDt.  
The mistag rate dependence originates from the kinematics of the physics sources for
wrong-charge kaons.
The three major sources of mistagged events in the 
\kaon\ category are wrong-sign \dz\ mesons from $B$ decays to
double charm, 
wrong-sign kaons from $D^+$ decays,
and kaons produced directly in $B$ decays.
All these sources produce a spectrum of charged tracks that have smaller 
$\sqrt{\Sigma p_t^2}$ than $B$ decays that produce a correct tag.
The \sigmaDt\ dependence originates from the 
$1/p^2_t$ dependence of \sigmaz\ for the individual contributing tracks.


\section{\Dt\ models for backgrounds}

\label{sec:bkgndmodel}


Although the true \Dt\ and resolution on \Dt\ are not well-defined  
for background events, we still
describe the total \Dt\ model as a ``physics model''
convolved with a ``resolution function''.

The background \Dt\ physics models we use in this analysis are
each a linear combination of one or more of the following terms,
corresponding to prompt (zero lifetime), exponential 
lifetime, and oscillatory 
distributions:
\begin{eqnarray*}
{\cal G}^\mathrm{pmt}_\mathrm{phys}(\Dttrue,g) & = &
{1\over 2}\delta(\Dttrue)\cdot\bigl( 1 + g\cdot (1- 
\omega^\mathrm{pmt})\bigr) \, ,\\
{\cal G}^\mathrm{life}_\mathrm{phys}(\Dttrue,g) & = & 
{1\over 4\tau^\mathrm{bg}}
\exp({-|\Dttrue|/\tau^\mathrm{bg}})\cdot\bigl(1 +
g\cdot(1-\omega^\mathrm{life})\bigr) \, ,\\ 
{\cal G}^\mathrm{osc}_\mathrm{phys}(\Dttrue,g) & = & 
{1\over 4\tau^\mathrm{bg}}
\exp({-|\Dttrue|/\tau^\mathrm{bg}})\cdot \bigl(1 +
g\cdot(1-\omega^\mathrm{osc})\cos\Delta m^\mathrm{bg} \Dttrue \bigr) \, ,\\ 
\end{eqnarray*}
where $\delta(\Dt)$ is a $\delta$-function, 
$g=+1$ for unmixed and $-1$ for mixed events, 
and $\tau^\mathrm{bg}$ and $\Delta m^\mathrm{bg}$ are the
effective lifetime and mixing frequency for the particular background.

For backgrounds, we use a resolution function that is the sum of
a narrow and a wide Gaussian distribution:
\begin{align*}
& {\cal
R}_\mathrm{G+G}(\dDt,\sigmaDt;b,s,f,b^w,s^w) \\ & =
f\cdot G(\dDt; b\sigmaDt,s\sigmaDt) + (1-f)\cdot
G(\dDt; b^w\sigmaDt,s^w\sigmaDt)\; .\\
\end{align*}
 
\subsection{Combinatoric background}
\label{sec:combbkgnd}

Events in which the \dst\ candidate corresponds to a random combination
of charged tracks (called combinatoric background) constitute the largest
background in the signal sample.
We use two sets of events to determine the appropriate parameterization of 
the \Dt\ model for combinatoric background:
events in data that are in the upper \massdiff\ sideband (above the peak due to real \dst\ decays); and
events in Monte Carlo simulation that are identified as combinatoric background,
based on the true information for the event, in both the \massdiff\ sideband and
peak region. 
The data and Monte Carlo \Dt\ distributions are described well by a  prompt plus
oscillatory term convolved with a double-Gaussian resolution function:
\begin{equation}
\begin{split}
{\cal G}^\mathrm{comb} = & \left[ f^\mathrm{osc}\cdot 
{\cal G}\mathrm{_{phys}^{osc}}
(\Dttrue,g;\tau^\mathrm{comb},\Delta
m^\mathrm{comb},\omega^\mathrm{osc})
+ 
(1- f^\mathrm{osc})\cdot
{\cal G}\mathrm{_{phys}^{pmt}}
(\Dttrue,g;\omega^\mathrm{pmt}) \right] \otimes \\
& {\cal R}_\mathrm{G+G}(\dDt,\sigmaDt;b,s,f,b^w,s^w) \;.
\label{eq:cb-mix-3}
\end{split}
\end{equation}

The parameters $\omega^\mathrm{pmt}$, $\Delta m^\mathrm{comb}$, $\tau^\mathrm{comb}$, 
$f$, $b^w$, and $s^w$ are shared among all control samples.
The parameters $\omega^\mathrm{osc}$, $f^\mathrm{osc}$, $b$, and $s$ 
are allowed
to be different depending on criteria such as tagging category, 
whether the data was recorded on- or off-resonance,
whether the candidate lepton passes real- or fake-lepton criteria,
whether the event passes the criteria for same-side or opposite-side \dst\ and $\ell$,
and how many identified leptons are in the event.
The total number of free parameters in the combinatoric background \Dt\ model is 24.

The relative fraction of $\Bz\Bzbar$ and $B^+B^-$ events in the combinatoric
background depends slightly on \massdiff. 
However, no significant dependence of the parameters
of the \Dt\ model on \massdiff\ is observed in data or Monte Carlo simulation.
The sample of events in the \massdiff\ sideband is used to determine the starting
values for the parameters in the final full fit to all data samples.

To reduce the total number of free parameters in the fit, parameters 
that describe 
the shape of the wide Gaussian (bias and width) 
are shared between combinatoric 
background and the three types of peaking background: 
continuum, fake-lepton, and uncorrelated-lepton.
The wide Gaussian fraction is allowed to be different for each type of
background. 

\subsection{Continuum peaking background}
\label{sec:contbkgnd}

All $c\overline c$ events that have a correctly
reconstructed \dst\ are defined as continuum peaking background,
independent of whether the associated lepton 
candidate is a real lepton or a fake lepton.
The $c\overline c$ Monte Carlo sample and off-resonance data are 
used to identify the appropriate \Dt\ model and sharing of parameters among 
subsamples.
The combinatoric-background \Dt\ model and parameters described in the 
previous section are used to model the combinatoric-background contribution in the
off-resonance \Dt\ distribution in data.

Events with a real \dst\ from
continuum $c\overline c$ production should have vanishing \Dt\ in
the case of perfect reconstruction. 
Therefore, we use the following model for the \Dt\ distribution of these
events:
$$
{\cal G}^\mathrm{cont} ={\cal
G}^\mathrm{pmt}_\mathrm{phys}(\Dttrue,g;\omega^\mathrm{pmt}) \otimes
{\cal R}_\mathrm{G+G}(\dDt,\sigmaDt;b,s,f,b^w,s^w) \; .
$$
Dependence on the flavor tagging information is included to accomodate
any differences in the amount of background events classified as mixed and unmixed.

By fitting to the data and Monte Carlo control samples
with different sharing of parameters across subsets of the data, 
we find that the apparent ``mistag
fraction'' for events  in the \kaon\ category is significantly different from the
mistag fraction for other tagging categories.
We also find that the core Gaussian bias 
is significantly different for opposite-side and same-side events.
We introduce separate parameters to accommodate these effects.

The total number of parameters used to describe the \Dt\ distribution of 
continuum peaking background is six.
The off-resonance control samples in data are  
used to determine starting values for the final full fit to all data samples.

\subsection{Fake-lepton peaking background}
\label{sec:fakebkgnd}

To determine the \Dt\ model and sharing of parameters for the fake-lepton 
peaking backgrounds, we use $\Bz\Bzbar$ and $B^+B^-$ Monte Carlo events in which
the \dst\ is correctly reconstructed 
but the lepton candidate is misidentified.
In addition, we use the fake-lepton control sample in data. 
The combinatoric and continuum peaking background \Dt\ models and parameters described
in the  previous two sections are used to model their
contribution to the fake-lepton \Dt\ distribution in data.
For this study, the contribution of signal is described by the signal parameters found
for signal events in the Monte Carlo simulation.

Since the fake-lepton peaking background is due to $B$ decays in which
the fake lepton and the \dst\ candidate can originate from the same $B$ or 
different $B$ mesons, 
we include both 
prompt and oscillatory terms in the \Dt\ model:
$$
{\cal G}^\mathrm{fake} = \left[ f^\mathrm{osc}\cdot {\cal
G}^\mathrm{osc}_\mathrm{phys} + (1-f^\mathrm{osc})\cdot {\cal
G}^\mathrm{pmt}_\mathrm{phys}\right] \otimes {\cal
R}_\mathrm{G+G}(\dDt,\sigmaDt;b,s,f,b^w,s^w) \; .
$$
We find that the apparent mistag rates for both the prompt and mixing terms, and the
bias of the core Gaussian of the resolution function, are different between some tagging
categories. 
The total number of parameters used to describe the fake-lepton background is 14.
The fake-lepton control samples in data are  
used to determine starting values for the final full fit to all data samples.

\subsection{Uncorrelated-lepton peaking background}
\label{sec:uncorrbkgnd}

To determine the \Dt\ model and sharing of parameters for the uncorrelated-lepton 
peaking backgrounds, we use $\Bz\Bzbar$ and $B^+B^-$ Monte Carlo events in which
the \dst\ is correctly reconstructed but the lepton candidate is from the other
$B$ in the event or from a secondary decay of the same $B$.
In addition, we use the same-side control sample in data, 
which is only about 30\%
uncorrelated-lepton background 
in the \massdiff\ peak region due to significant contributions from 
combinatoric  background
and signal.
The combinatoric and other peaking background \Dt\ models and parameters described
in the  previous two sections are used to model their
contribution to the same-side \Dt\ distribution in data.
For this initial study, the contribution of signal is described by the signal
parameters found for signal events in the Monte Carlo simulation.

Physics and vertex reconstruction considerations suggest several features of the
\Dt\ distribution for the uncorrelated-lepton sample.  
First, we expect the reconstructed \Dt\ to be systematically
smaller than the true \Dt\ value since using a lepton and a \dst\ from 
different $B$ decays  
will generally reduce the separation between the reconstructed
\bdstl\ and \btag\ vertices. 
We also expect that events with small true \Dt\ will have a higher
probability of being misreconstructed as an uncorrelated lepton
candidate because it is more likely that the fit of the \dst\ and $\ell$ to
a common vertex will converge for these events.
Finally, we expect truly mixed events to have a higher fraction of uncorrelated-lepton
events because in mixed events the charge of the \dst\ is opposite that of
primary leptons on the tagging side.
These expectations are confirmed in the Monte Carlo simulation.

We do not expect the uncorrelated-lepton background to exhibit any mixing behavior
and none is observed in the data or Monte Carlo control samples.
We describe the \Dt\ distribution with the sum of a lifetime term and a prompt term, 
convolved with a double-Gaussian resolution function:
\begin{align}
{\cal G}^\mathrm{uncor} = & \left[ f^\mathrm{life}\cdot 
{\cal G}\mathrm{_{phys}^{life}}
(\Dttrue,g;\tau^\mathrm{uncor},
\omega^\mathrm{life})
+ 
(1- f^\mathrm{life})\cdot
{\cal G}\mathrm{_{phys}^{pmt}}
(\Dttrue,g;\omega^\mathrm{pmt}) 
\right] \otimes  \notag \\
& {\cal R}_\mathrm{G+G}(\dDt,\sigmaDt;b,s,f,b^w,s^w) \;.
\label{eq:unco}
\end{align}
The effective mistag rates $\omega^\mathrm{pmt}$ and $\omega^\mathrm{life}$
accommodate different fractions
of uncorrelated-lepton backgrounds in events classified as mixed and unmixed.
We find that the apparent mistag
rate for the lifetime term is different between some tagging categories. 
All other parameters are consistent among the different subsamples. 
The total number of parameters used to describe the uncorrelated-lepton background is six.
The uncorrelated-lepton control samples in data are  
used to determine starting values for the final full fit to all data samples.

\subsection{Charged $B$ peaking background}
\label{sec:chBbkgnd}

The charged-$B$ peaking background is due to decays of the type 
$B^\pm\rightarrow D^*\ell\nu_\ell X$.
Since charged $B$'s do not exhibit mixing behavior,
our strategy is to use the \Dt\ and tagging information
to discriminate charged-$B$ peaking background events
from neutral-$B$ signal events, in the simultaneous fit to all samples. 
We use the same resolution model and parameters as for the  
neutral-$B$ signal since the decay
dynamics are very similar. 
The signal model, with the charged $B$ background
term,  becomes
\begin{equation*}
\begin{split}
{\cal G}^{\mathrm{sig}}(\Dt,\sigmaDt,g\,;\vec{q}_c^{\, \mathrm{sig}}) 
& = [ \frac{1-\fBp}{4\tauBz} e^{-|\Dttrue|/\tauBz}\left(1+ g (1-2 \omega_{\Bz}^c)\cos(\Dm\Dttrue)\right) +
\\
 & \quad \frac{\fBp}{4\tauBp} e^{-|\Dttrue|/\tauBp}\left(1+ g (1-2 \omega_{B^+}^c)
\right)] \otimes {\cal R}(\dDt,\sigmaDt; \vec{q}_c) \, ,\\
\end{split}
\end{equation*}
where $\omega_{\Bz}^c$ ($\omega_{B^+}^c$) is the mistag fraction for
neutral (charged) $B$ mesons for tagging category  $c$.

Given that the ratio of the charged $B$ to neutral $B$ lifetime is close to
1 and the fraction of charged $B$ mesons in the peaking sample is small, we
do not have sufficient sensitivity to distinguish the lifetimes in the fit. 
We parameterize the physics model for the $B^+$ in terms of the 
lifetime ratio 
$\tauBp/\tauBz$, and fix this ratio to the Review of Particle Properties 2002 world average \cite{ref:PDG2002}.
We vary the ratio by the error on the world average to 
estimate the corresponding systematic
uncertainties on \tauBz\ and \Dm\ (see Sec.~\ref{sec:systematics}).

The fit is sensitive to only two parameters 
among \wBp, \wBz\ and the charged $B$ fraction (\fBp). 
Therefore we fix the ratio  of mistag rates, $\wBp/\wBz$,
to the value of the ratio measured with fully reconstructed charged and 
neutral hadronic $B$ decays in data, for each tagging category.


\section{Fit results}

\label{sec:results}

The total number of free parameters in the final fit is 72: 
22 in the signal model, 
24 in the combinatoric background model, 
and 26 in peaking background models. 
The fitted signal \Dt\ model parameters are shown in
Table~\ref{tab:result-signal}. 

\begin{table}[htb]
\caption{Results of full fit to data --- signal model and 
resolution function parameters.
A small correction, described in Sec.~\ref{sec:MCtests}, has been
applied to \tauBz\ and \Dm.}
\begin{center}
\begin{tabular}{|lr|lr|lr|}
\hline
\multicolumn{6}{|c|}{Signal Model and \Dt\ Resolution Function Parameters} \\
\hline\hline
parameter & value & parameter & value & parameter & value \\
\hline
\Dm (ps$^{-1}$) & $ 0.492\pm 0.018$ & \fBp & $ 0.082\pm 0.029$ & $s$ & $ 1.201\pm 0.063$ \\ 
\tauBz (ps) & $ 1.523^{+0.024}_{-0.023}$ & $\omega_\mathrm{lepton}$ & $ 0.071\pm 0.015$ & $\kappa$ & $ 0.86\pm 0.17$ \\ 
- & - & $\omega\mathrm{_{kaon}^{offset}}$ & $ 0.002\pm 0.024$ & $f\mathrm{_{lepton}}$ & $ 0.72\pm 0.10$ \\ 
- & - & $m_\mathrm{kaon}$ & $ 0.229\pm 0.036$ & $f_\mathrm{kaon}$ & $ 0.609\pm 0.088$ \\ 
- & - & $\omega_\mathrm{NT1}$ & $ 0.212\pm 0.020$ & $f\mathrm{_{NT1}}$ & $ 0.69\pm 0.13$ \\ 
- & - & $\omega_\mathrm{NT2}$ & $ 0.384\pm 0.018$ & $f\mathrm{_{NT2}}$ & $ 0.70\pm 0.10$ \\ 
- & - & $\omega_\mathrm{NT3}$ & $ 0.456\pm 0.012$ & $f\mathrm{_{NT3}}$ & $ 0.723\pm 0.078$ \\ 
- & - & $\Delta \omega_\mathrm{lepton}$ & $-0.001\pm 0.022$ & $f\mathrm{^{out}}$ & $ 0.0027\pm 0.0017$ \\ 
- & - & $\Delta \omega_\mathrm{kaon}$ & $-0.024\pm 0.015$ & $b\mathrm{^{out}}$ (ps) & $-5.000$ \\ 
- & - & $\Delta \omega_\mathrm{NT1}$ & $-0.098\pm 0.032$ & $s\mathrm{^{out}}$ (ps) & $ 6.000$ \\ 
- & - & $\Delta \omega_\mathrm{NT2}$ & $-0.112\pm 0.028$ & - & - \\ 
- & - & $\Delta \omega_\mathrm{NT3}$ & $-0.023\pm 0.019$ & - & - \\ 
\hline\hline
\end{tabular}
\end{center}
\label{tab:result-signal}
\end{table}

The statistical correlation coefficient 
between \tauBz\ and \Dm\ 
is $\rho(\Dm,\tauBz)= -0.22$. 
The global correlation coefficients for \tauBz\ and \Dm, and
some of the correlation
coefficients between
\tauBz\ or \Dm\ and other parameters, are 
shown in Table~\ref{tab:result-cor}.

\begin{table}[htb]
\caption{Global correlation coefficients for 
\Dm\ and \tauBz\ from the full fit to data and other
correlation coefficients for pairs of key parameters
in the fit.}
\begin{center}
\begin{tabular}{lr}
\hline\hline
\Dm\ global correlation &  0.74 \\
\tauBz\ global correlation &  0.69 \\
\hline
$\rho(\Dm,\tauBz)$ & $-0.22$ \\
\hline
$\rho(\Dm,\fBp)$ & 0.58 \\
$\rho(\tauBz,\sigma\mathrm{^1_{sig}})$ & $-0.49$ \\
$\rho(\tauBz,f\mathrm{_{sig}^{out}})$ & $-0.26$ \\
\hline\hline
\end{tabular}
\end{center}
\label{tab:result-cor}
\end{table}

\begin{figure}[htb]
\begin{center}
\includegraphics[width=3in]{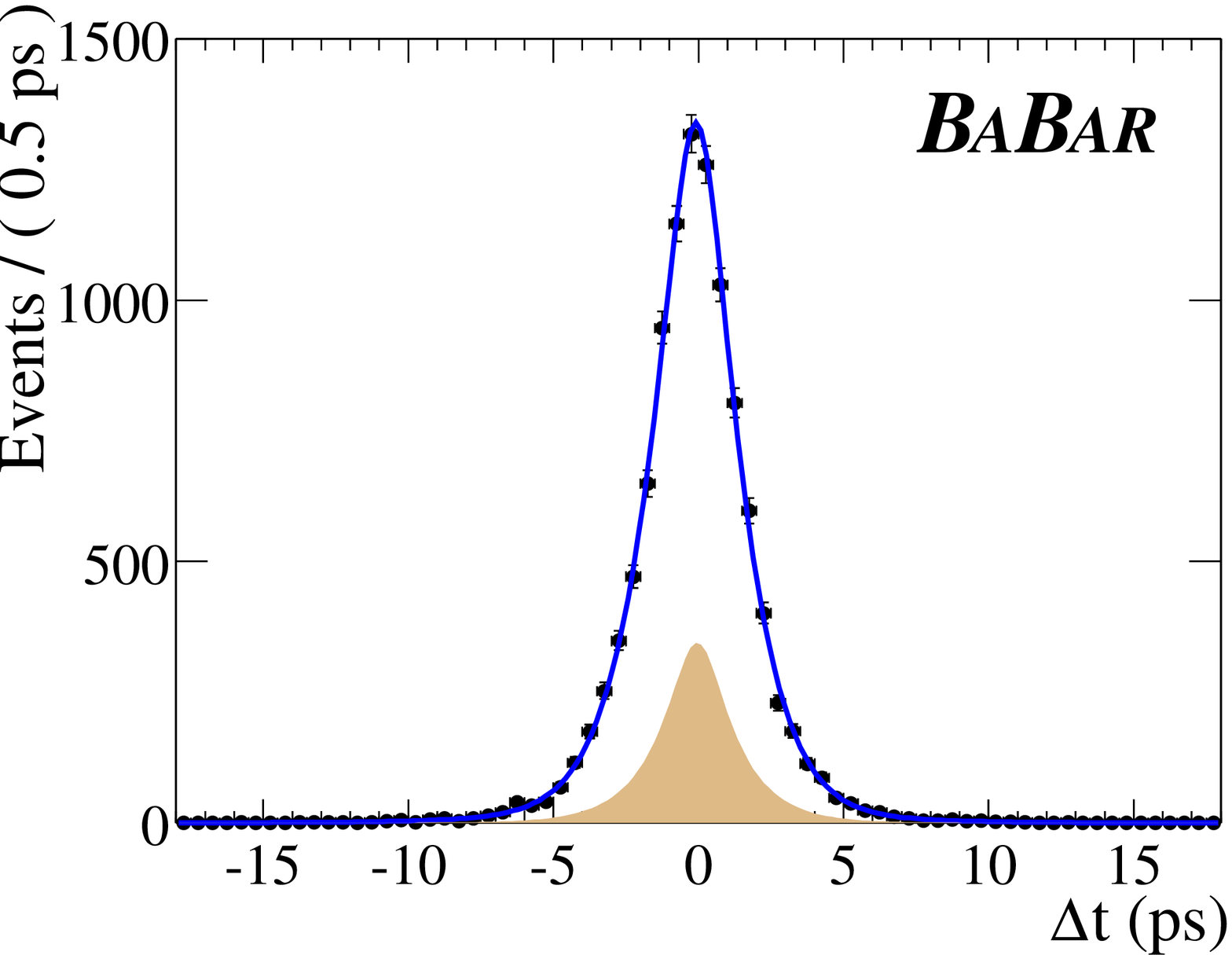}
\includegraphics[width=3in]{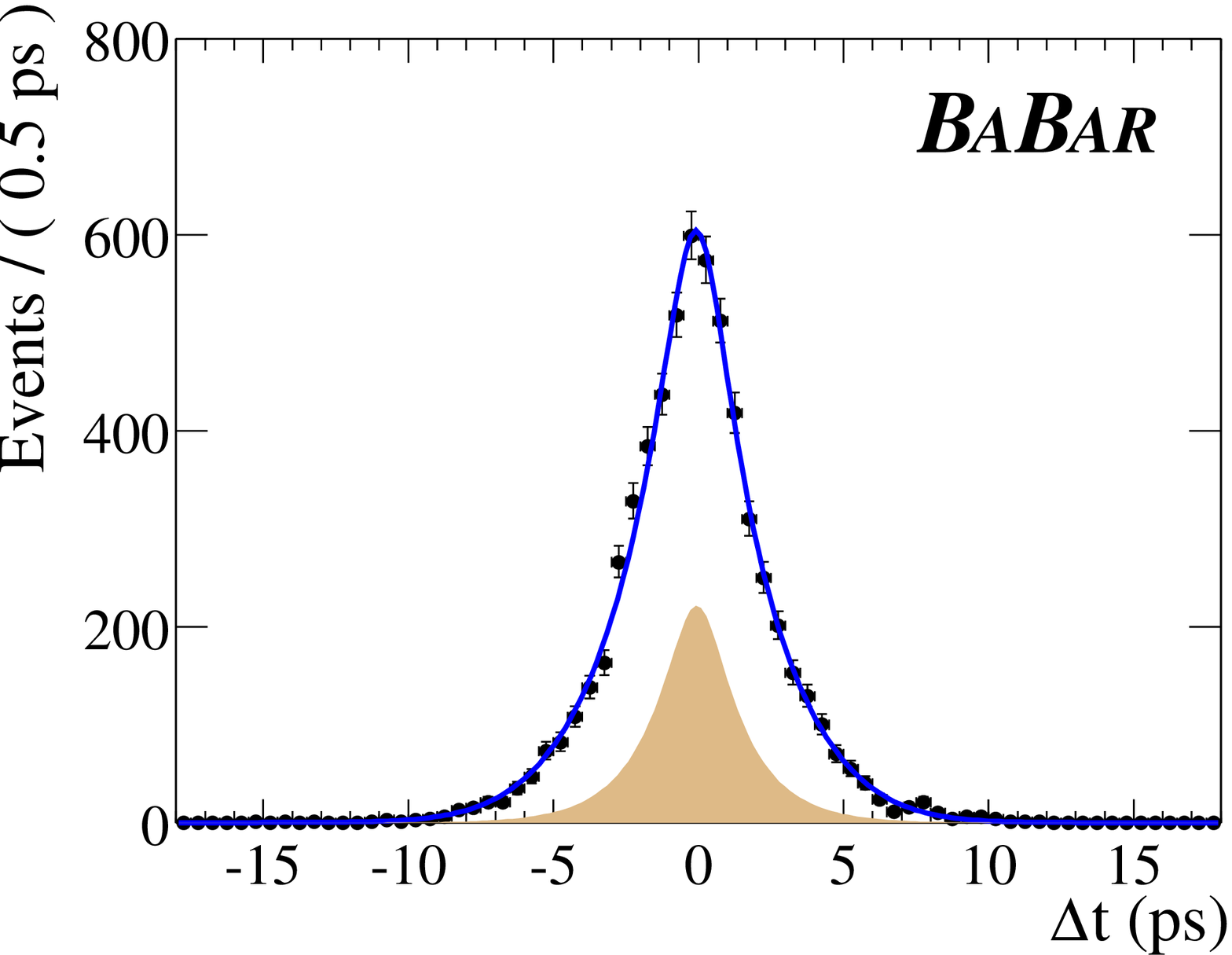} \\
\includegraphics[width=3in]{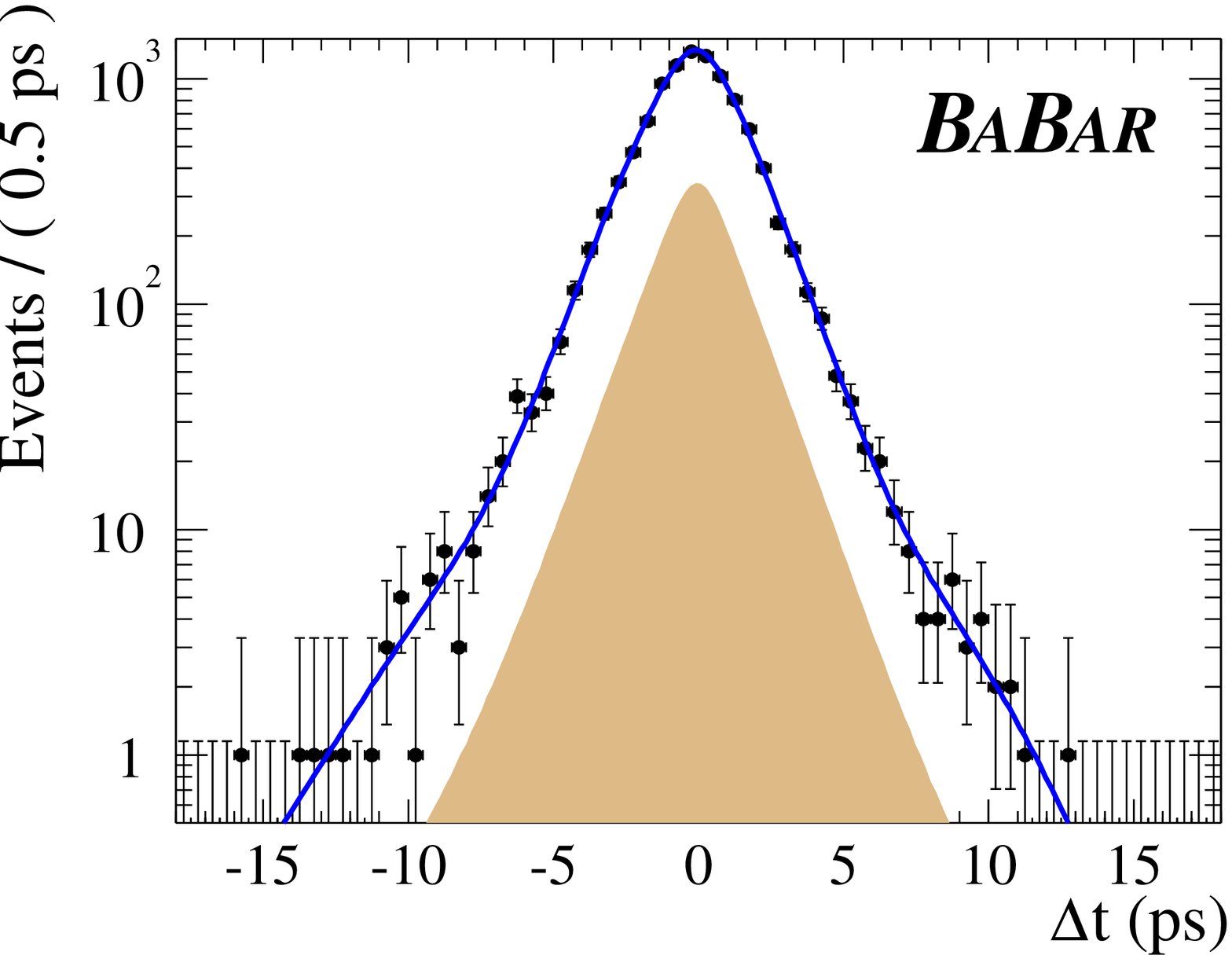}
\includegraphics[width=3in]{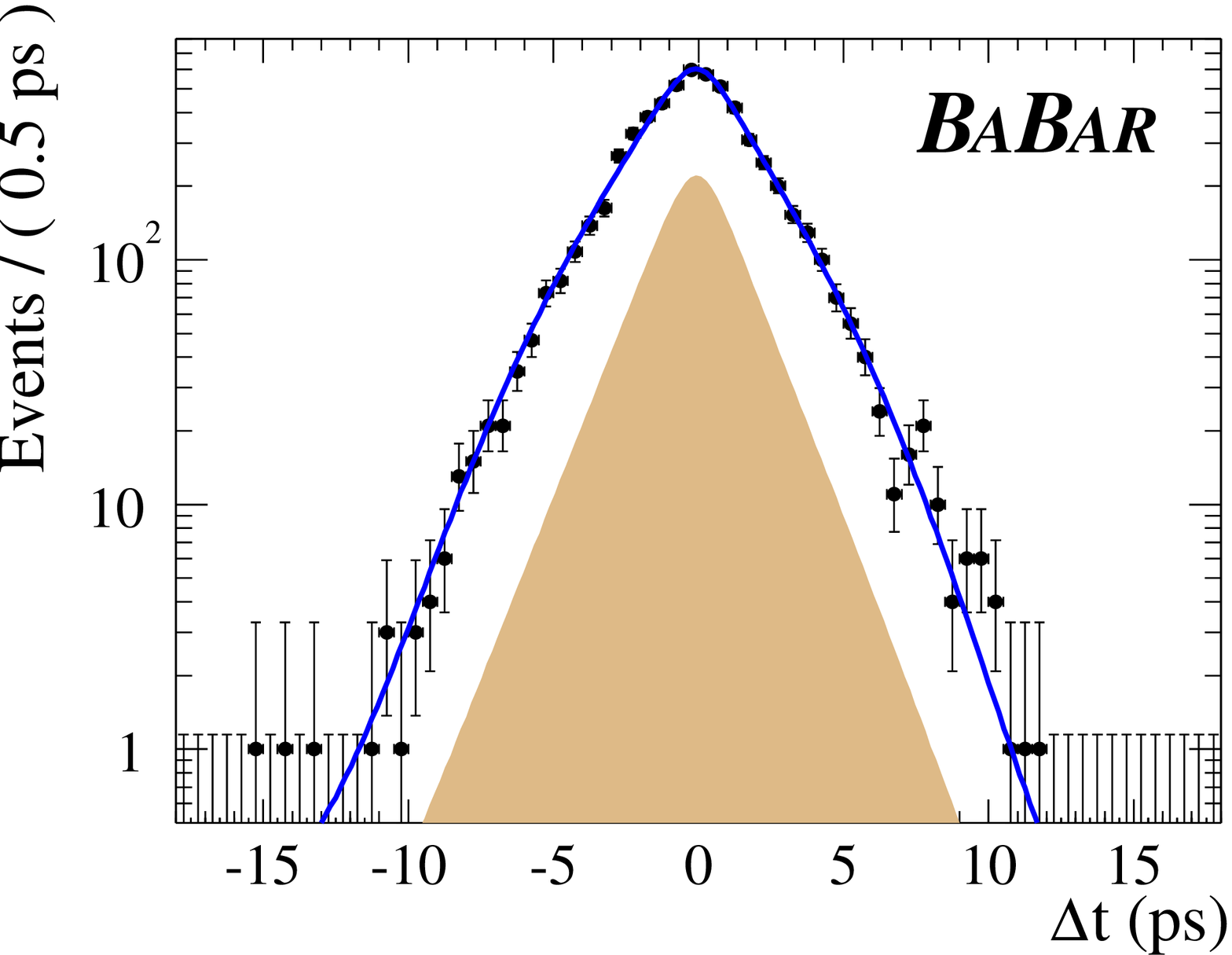}
\end{center}
\caption{The \Dt\ distribution for unmixed and mixed events in the signal
sample
(opposite-side \dst-lepton candidates in on-resonance data)
and the projection of the fit results.
The left-hand plots are for unmixed events; the right-hand plots 
for mixed events.
The shaded area
shows the background contribution to the distributions.}
\label{fig:result-dt}
\end{figure}

\begin{figure}[htb]
\begin{center}
\includegraphics[width=4in]{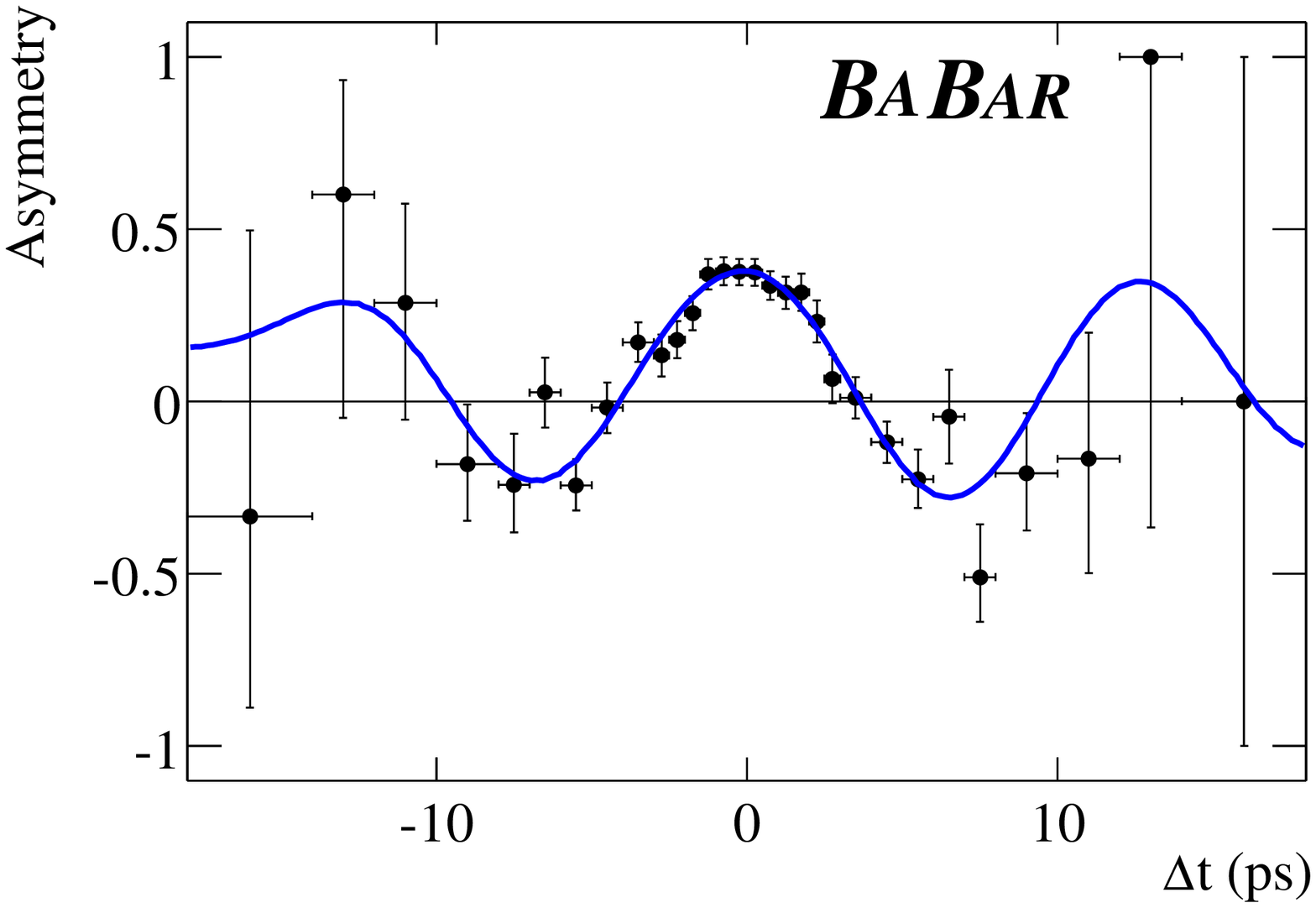}
\end{center}
\caption{The asymmetry plot for mixed and unmixed events in
the signal sample (opposite-side \dst-lepton candidates in 
on-resonance data) in the \dm\ range from 143 to 148~MeV,
and the projection of the fit results. Errors on the data points
are computed by considering the binomial probabilities for 
observing different numbers of mixed and unmixed
events while preserving the total number. }
\label{fig:result-asym}
\end{figure}

Figure~\ref{fig:result-dt} 
shows the \Dt\ distributions for unmixed and mixed events in 
the signal sample 
(opposite-side \dst-lepton candidates in on-resonance data).
The points correspond to data.
The curves correspond to the sum of the
projections of the appropriate relative amounts of signal and 
background \Dt\ models for the signal sample in the \dm\ 
range between 143 and 148~MeV.
Figure~\ref{fig:result-asym} shows the asymmetry
$$A =   {N_\mathrm{unmixed}(\Dt) - N_\mathrm{mixed}(\Dt)
  \over  N_\mathrm{unmixed}(\Dt) + N_\mathrm{mixed}(\Dt)}.$$
The unit amplitude for the cosine dependence of 
$A$ is diluted by the mistag probability, the experimental
\Dt\ resolution, and backgrounds.

\begin{figure}[htb]
\begin{center}
\includegraphics[width=4in]{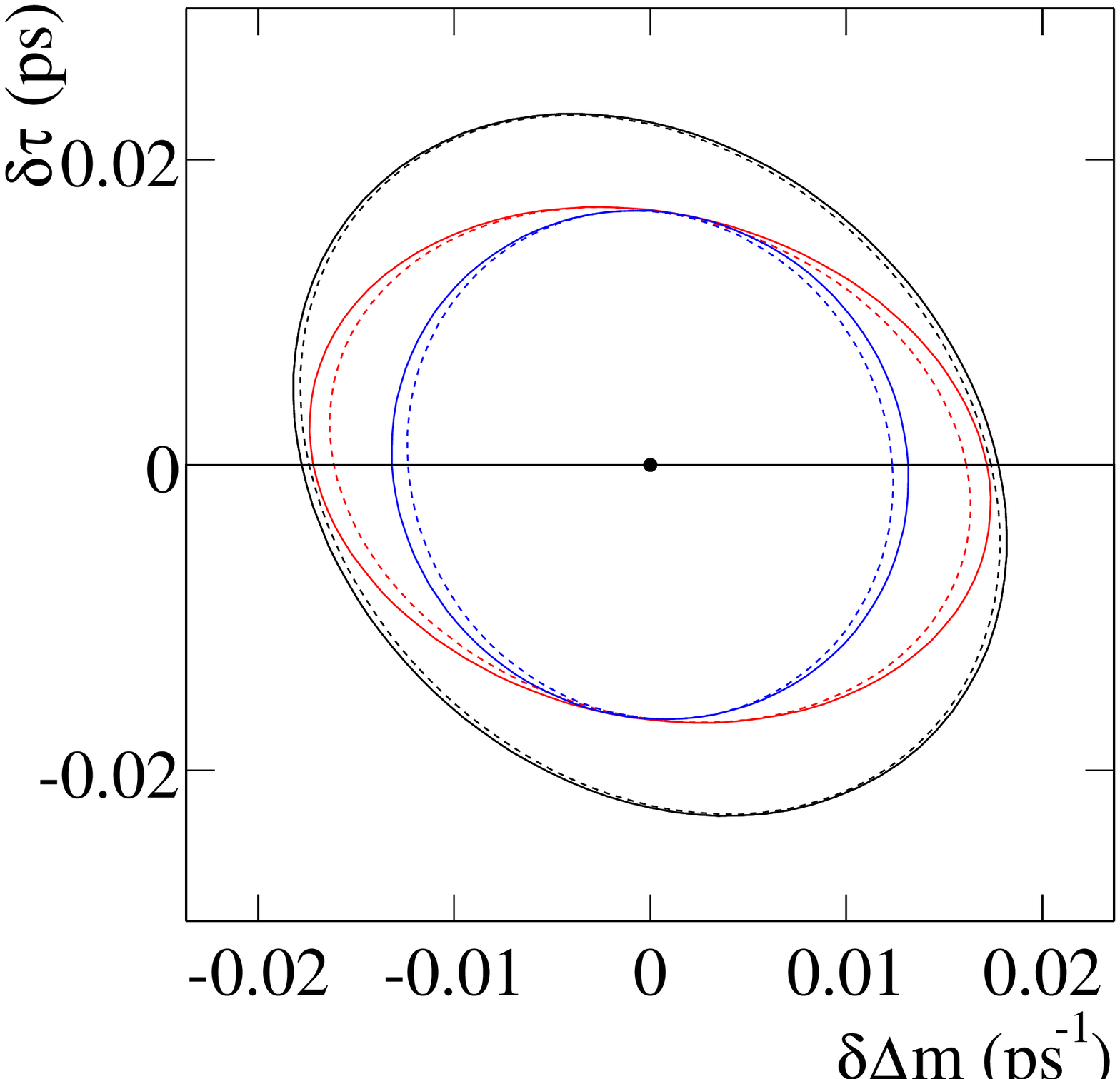}
\end{center}
\caption{Comparison of one-sigma error ellipses in the  \Dm-\tauBz\
plane for fits in which different sets of parameters are free. From
the innermost to the outermost ellipse, the floating parameters are 
(\Dm, \tauBz),
(\Dm, \tauBz, mistag fractions), (\Dm, \tauBz, \fBp), (\Dm, \tauBz,
\fBp, mistag fractions), all
signal \Dt\ parameters, and the default fit (72 floating parameters).}
\label{fig:contourNest}
\end{figure}

Since we float many parameters in the model, it is interesting to see
how the errors on \tauBz\ and \Dm, and their correlation change when 
different parameters are free in the fit, 
or fixed to their best value from the full fit.
We perform a series of fits, fixing all
parameters at the values obtained from the default fit, except (a) \Dm\
and \tauBz, (b) \Dm, \tauBz, and all mistag fractions in the signal model, 
(c) \Dm, \tauBz, and \fBp, (d) \Dm, \tauBz, \fBp, and all mistag fractions
in the signal model, (e) all parameters in the signal \Dt\ model. 
The one-sigma error ellipses for these fits
and for the default fit are shown in Fig.~\ref{fig:contourNest}.

We can see that the error on \tauBz\ changes very little until we float
the signal resolution function. Floating the background parameters adds
a very small contribution to the error. 
The contribution from the charged $B$
fraction and mistag fractions to the \tauBz\ error is negligible. 
On the other hand, the charged $B$ fraction changes the error on \Dm\ the
most. The contributions from floating the mistag fractions, resolution
functions, and background \Dt\ models are relatively small.

\section{Validation and cross checks}

\label{sec:validations}


In Sec.~\ref{sec:MCtests}, we describe several tests of the fitting 
procedure that were performed with both fast parameterized Monte Carlo
simulations and full detector simulations.
In Sec.~\ref{sec:datachecks}, we give the results of performing
cross-checks on data, including fitting to different subsamples of the
data and fitting with variations to the standard fit.

\subsection{Tests of fitting procedure with Monte Carlo simulations}
\label{sec:MCtests}

A test of the fitting procedure is performed with fast 
parameterized Monte Carlo simulations, where 87 experiments are 
generated with signal and background control sample sizes and
compositions corresponding to that obtained from the full likelihood fit
to data.   
The mistag rates and \Dt\ distributions are generated
according to the model used in the likelihood fit.
The full fit is then performed on each of these experiments.
We find no statistically significant bias in
the average values of \tauBz\ and \Dm\ for the 87 fits.
The RMS spread in the distribution of results is consistent with
the mean statistical error from the fits and the statistical
error on the results in data, for both \tauBz\ and \Dm. 
We find that 17 of the fits to the 87 experiments result in a value
of the negative log likelihood that is smaller (better) than that found
in data. 
We also check the statistical errors on data 
by measuring the increase in negative log likelihood 
in data in the two-dimensional (\tauBz, \Dm) space in the vicinity of the 
minimum of the negative log likelihood.  
We found that the positive error on \tauBz\ is about 6\% larger than 
that predicted by the fitting program, whereas the other errors are 
the same as predicted.
We increased the positive statistical error on \tauBz\ by 6\%.

We also fit two types of Monte Carlo samples that include full detector
simulation: pure signal and signal plus background.
To check whether the selection criteria introduce any bias in the 
lifetime or mixing frequency,  we fit the signal physics model to
the true lifetime distribution, using true tagging information, for
a large sample of signal Monte Carlo events that pass all selection
criteria. 
We also fit the measured \Dt\ distribution, using measured tagging 
information, with the complete signal \Dt\ model described in 
Sec.~\ref{sec:sigmodel}.
We find no statistically significant bias in the values of \tauBz\ or 
\Dm\ extracted in these fits.

The \BzBzbar, $B^+B^-$, and $c\overline c$ Monte Carlo samples that
provide simulated background events along with signal events are much
smaller than the pure signal Monte Carlo samples.
In addition, they are not much larger than the data samples.  
In order to increase the statistical sensitivity to any bias introduced
when the  background samples are added to the fit, we compare the 
values of \tauBz\ and \Dm\ from the fit to signal plus background
events, and pure signal events from the same sample.
We find that when background is added, the value of \tauBz\ increases by
$(0.022 \pm 0.009)$~ps and the value of \Dm\ increases by 
$(0.020 \pm 0.005)$~ps$^{-1}$, 
where the 
error is the difference in quadrature between the statistical errors
from the fit
with and without background.
We correct our final results in data for these biases, which are each
roughly the same size as the statistical error on the results in data.
We conservatively apply a systematic uncertainty on this  bias
equal to the full statistical error on the measured result in Monte
Carlo simulation with background:  $\pm 0.018$~ps for \tauBz\ 
and $\pm 0.012$~ps$^{-1}$ for 
\Dm. 

\subsection{Cross-checks in data}
\label{sec:datachecks}

We perform the full maximum-likelihood fit on different subsets
of the data and find no statistically significant difference in the
results for different subsets.
The fit is performed on datasets divided according to tagging category, 
$b$-quark flavor of the \dstl\ candidate, $b$-quark flavor of the tagging
$B$, and $D^0$ decay mode.
We also vary the range of \Dt\ over which we perform the fit 
(maximum value of $|\Dt|$ equal to 10, 14, and 18~ps), 
and decrease the maximum allowed value of \sigmaDt\ from 1.8~ps to 1.4~ps.
Again, we do not find statisitically significant changes in the values of 
\tauBz\ or \Dm.


\section{Systematic studies}

\label{sec:systematics}


We estimate systematic uncertainties on the parameters \tauBz\ and
\Dm\ with studies performed on both data and Monte Carlo samples, and
obtain the results summarized in Table~\ref{tab:systematics}. 

\begin{table}[htb]
\caption{Summary of systematic uncertainties on the two physics parameters, 
\tauBz and \Dm.}
\begin{center}
\begin{tabular}{|l|rr|c|}
\hline
Source                & $\delta(\Dm)$ (ps$^{-1}$) & $\delta(\tauBz)$ (ps) \\
\hline\hline
Selection and fit bias        & $0.0123$ & $0.0178$ \\
$z$ scale                     & $0.0020$ & $0.0060$ \\
PEP-II boost                  & $0.0005$ & $0.0015$ \\
Beam spot position            & $0.0010$ & $0.0050$ \\
SVT alignment                 & $0.0030$ & $0.0056$ \\
Background / signal prob.     & $0.0029$ & $0.0032$ \\
Background $\Delta t$ models  & $0.0012$ & $0.0063$ \\
Fixed $B^+$/$B^0$ lifetime ratio & $0.0003$ & $0.0019$ \\
Fixed $B^+$/$B^0$ mistag ratio    & $0.0001$ & $0.0003$ \\
Fixed signal outlier shape    & $0.0010$ & $0.0054$ \\
Signal resolution model       & $0.0009$ & $0.0034$ \\
\hline
Total systematic error             & $0.013\ $ & $0.022\ $   \\
\hline
\end{tabular}
\end{center}
\label{tab:systematics}
\end{table}

The largest source of systematic uncertainty on both parameters is the
limited statistical precision for determining the bias due to the
fit procedure (in particular, the background modeling) with Monte
Carlo events. We assign the statistical errors of a full fit to Monte
Carlo samples including background to estimate this systematic uncertainty.
See Sec.~\ref{sec:MCtests} for more details.

The calculation of a decay-time difference \Dt\ for each event
assumes a nominal detector $z$-scale, PEP-II boost, vertical beam-spot position, and SVT
internal alignment. We vary each of
these assumptions and assign the variation in the fitted parameters as
a corresponding systematic uncertainty.

The  modeling of the background contributions to the sample determines
the probability we assign for each event to be due to signal and the
\Dt\ distribution we expect for background events. We estimate the
systematic uncertainty due to the assumed background \Dt\
distributions as the shift in the fitted parameters when we
replace the model for the largest
background (due to combinatoric events) with a pure lifetime model.
We estimate the uncertainty due to the signal
probability calculations by repeating the full fit using an ensemble
of different signal and background parameters for the \massdiff\
distributions, varied randomly 
according to the measured statistical uncertainties and  correlations 
between the parameters. We assign
the spread in each of the resulting fitted physics parameter as the systematic
uncertainty.

The model of the charged $B$ background assumes fixed $B^+/\Bz$ ratios
for the mistag rates and lifetimes. 
We vary the mistag ratio by the uncertainty determined from
separate fits to hadronic events.
We vary the lifetime ratio by the statistical uncertainty on the world 
average~\cite{ref:PDG2002}.
The resulting change in the fitted physics  parameters 
is assigned as a systematic uncertainty.

The final category of systematic uncertainties is due to assumptions
about the resolution model for signal events. We largely avoid
assumptions by floating many parameters to describe the resolution
simultaneously with the parameters of interest. However, two sources
of systematic uncertainty remain: the shape of the outlier
contribution, which cannot be determined from data alone, and the
assumed parameterization of the resolution for non-outlier
events. We study the sensitivity to the outlier shape by repeating the
full fit with an ensemble of different shapes, and assign the spread
of the resulting fitted values as a systematic uncertainty. We
estimate the uncertainty due to the assumed resolution
parameterization by repeating the full fit with a triple-Gaussian
resolution model and assigning the shift in the fitted values as the
uncertainty.

The total systematic uncertainty on \tauBz\ is 0.022~ps and on
\Dm\ is 0.013~ps$^{-1}$.


\section{Summary}

\label{sec:Summary}

We use a sample of approximately 14,000 exclusively reconstructed
$B^0 \rightarrow D^{*-}\ell^+\nu_\ell$ signal events to 
measure the $B^0$ lifetime $\tauBz$ and oscillation frequency $\Dm$
simultaneously, with an unbinned maximum-likelihood fit.
The preliminary results are 
$$\tauBz = (1.523^{+0.024}_{-0.023} \pm 0.022)~\rm{ps}$$
and
$$\Dm = (0.492 \pm 0.018 \pm 0.013)~\rm{ps}^{-1}.$$
The statistical correlation coefficient between \tauBz\ and \Dm\ is $-0.22$.
Both the lifetime and mixing frequency have combined statistical and 
systematic uncertainties that are comparable to those of the most
precise previously-published experimental measurements~\cite{ref:PDG2002}.
The results are consistent with the world average measurements of 
$\tauBz = (1.542\pm 0.016)~\rm{ps}$
and
$\Dm = (0.489 \pm 0.008)~\rm{ps}^{-1}$~\cite{ref:PDG2002}.


\section{Acknowledgments}

\label{sec:Acknowledgments}

We are grateful for the 
extraordinary contributions of our \pep2\ colleagues in
achieving the excellent luminosity and machine conditions
that have made this work possible.
The success of this project also relies critically on the 
expertise and dedication of the computing organizations that 
support \babar.
The collaborating institutions wish to thank 
SLAC for its support and the kind hospitality extended to them. 
This work is supported by the
US Department of Energy
and National Science Foundation, the
Natural Sciences and Engineering Research Council (Canada),
Institute of High Energy Physics (China), the
Commissariat \`a l'Energie Atomique and
Institut National de Physique Nucl\'eaire et de Physique des Particules
(France), the
Bundesministerium f\"ur Bildung und Forschung and
Deutsche Forschungsgemeinschaft
(Germany), the
Istituto Nazionale di Fisica Nucleare (Italy),
the Research Council of Norway, the
Ministry of Science and Technology of the Russian Federation, and the
Particle Physics and Astronomy Research Council (United Kingdom). 
Individuals have received support from 
the A. P. Sloan Foundation, 
the Research Corporation,
and the Alexander von Humboldt Foundation.


\end{document}